\documentclass[superscriptaddress,aps,pra,10pt]{revtex4-2}

\usepackage{chemformula} 
\usepackage{algorithm2e}
\usepackage{amsmath, amssymb}
\usepackage[hidelinks]{hyperref}

\usepackage{siunitx} 
\DeclareSIUnit\molar{\textsc{M}}
\DeclareSIUnit\adu{ADU}
\DeclareSIUnit\electron{e^-}
\DeclareSIUnit\photon{photon}
\DeclareSIUnit\mirror{mirror}
\DeclareSIUnit\frame{frame}
\DeclareSIUnit\rpm{rpm}
\DeclareSIUnit\unit{unit}
\DeclareMathOperator{\na}{NA}
\DeclareMathOperator{\argmin}{arg min}
\DeclareMathOperator{\prox}{prox}
\DeclareMathOperator{\qe}{QE}

\newcommand{\abs}[1]{\left\lvert #1 \right\rvert}
\renewcommand{\vec}[1]{\mathbf{#1}}
\newcommand{\uvec}[1]{\hat{\mathbf{#1}}}
\newcommand{\ft}[1]{\mathcal{F}_\perp \left \{#1  \right \}}
\newcommand{\ift}[1]{\mathcal{F}_\perp^{-1} \left \{#1 \right \}}

\newcommand{\videoTrackingBeads}{1}
\newcommand{\videoTrackingBeadsDense}{2}
\newcommand{\videoEcoli}{3}
\newcommand{\videoEcoliThreeD}{4}
\newcommand{\videoKhzBeads}{5}
\newcommand{\videoKhzBeadsDenser}{6}

\newcommand{\odtDetails}{\ref{section:odt_details}}
\newcommand{\dmdEff}{\ref{section:dmd_eff}}
\newcommand{\reconstruction}{\ref{section:reconstruction}} 
\newcommand{\linscatt}{\ref{section:linear_scattering}}

\newcommand{\mieRecon}{\ref{fig:mie_reconstruction}} 
\newcommand{\lossFnMix}{\ref{eq:loss_fn_mix}}
\newcommand{\rytovPhase}{S\ref{eq:rytov_phase}}

\begin{document}

\title{Fourier synthesis optical diffraction tomography for kilohertz rate volumetric imaging}

\author{Peter T. Brown}
\email{ptbrown2@asu.edu}
\affiliation{Center for Biological Physics and Department of Physics, Arizona State University, Tempe, AZ 85287, USA}
\author{Nikta Jabbarzadeh}
\affiliation{Center for Biological Physics and Department of Physics, Arizona State University, Tempe, AZ 85287, USA}
\author{Aidan Pintuff}
\affiliation{Biology Department, Western Washington University, Bellingham, WA 98225, USA}
\author{Luis Meneses}
\affiliation{Biodesign Center for Mechanisms of Evolution, Arizona State University, Tempe, AZ 85287, USA}
\author{Ekaterina Monakhova}
\affiliation{Biology Department, Western Washington University, Bellingham, WA 98225, USA}
\author{Rory Kruithoff}
\affiliation{Center for Biological Physics and Department of Physics, Arizona State University, Tempe, AZ 85287, USA}
\author{Navish Wadhwa}
\affiliation{Center for Biological Physics and Department of Physics, Arizona State University, Tempe, AZ 85287, USA}
\affiliation{Biodesign Center for Mechanisms of Evolution, Arizona State University, Tempe, AZ 85287, USA}
\author{Domenico F. Galati}
\affiliation{Biology Department, Western Washington University, Bellingham, WA 98225, USA}
\author{Douglas P. Shepherd}
\email{douglas.shepherd@asu.edu}
\affiliation{Center for Biological Physics and Department of Physics, Arizona State University, Tempe, AZ 85287, USA}

\begin{abstract}
Many biological and soft matter processes occur at high speeds in complex 3D environments, and developing imaging techniques capable of elucidating their dynamics is an outstanding experimental challenge. Here, we introduce Fourier Synthesis Optical Diffraction Tomography (FS-ODT), a novel approach for high-speed quantitative phase imaging capable of recording the 3D refractive index at kilohertz rates. FS-ODT introduces new pattern generation and inverse computational strategies that multiplex tens of illumination angles in a single tomogram, dramatically increasing the volumetric imaging rate. We validate FS-ODT performance by imaging samples of known composition and accurately recovering the refractive index for increasing pattern complexity. We further demonstrate the capabilities of FS-ODT for probing complex systems by studying the hindered diffusion of colloids in solution and the motility of single-cellular bacterial swimmers. We believe that FS-ODT is a promising approach for unlocking challenging imaging regimes in biophysics and soft matter that have been little explored, including understanding the physical interactions of colloids and microswimmers with their viscous 3D environment and the interplay between these stimuli and the molecular response of biological systems.
\end{abstract}

\maketitle

\section*{Introduction}
Probing biology and physics at high resolution and speed in 3D is a tremendous experimental challenge that has inspired significant advances in microscopy techniques across time and length scales. Fluorescence imaging is one of the most widely used techniques, with many existing variants addressing different experimental needs from the nano- to macro-scale. Fundamentally, all fluorescence methods are subject to the ``triangle of frustration", where photobleaching, phototoxicity and fluorophore photophysics limit the speed, duration, and total light dose~\cite{Laissue2017}. 

Recent advances in quantitative phase imaging (QPI) techniques have positioned these as an alternative to fluorescence in many domains~\cite{Javidi2021}. In contrast to fluorescence microscopy, image contrast in QPI originates from the sample's refractive index (RI), avoiding the aforementioned low-signal levels, photobleaching, and the triangle of frustration associated with fluorescent labeling. When morphology and dynamics are of primary interest, QPI approaches are becoming a method of choice for long-term, volumetric imaging at few-hertz rates~\cite{parkQuantitativePhaseImaging2018}. 

For acquisition rates beyond a few hertz, a variety of single-shot QPI approaches enable fast ``volumetric'' imaging, including digital in-line holography~\cite{Xu2001}, off-axis holography~\cite{girshovitzFastPhaseProcessing2015}, and iSCAT~\cite{Lindfors2004} (table~\ref{table:khz_methods}). However, single-shot QPI methods can only obtain unambiguous volumetric reconstructions by putting strong priors on the geometry of particles that are imaged, e.g., spheres, rods, or other particles with a known scattering model. Such approaches commonly fail in dense samples and have limited axial resolution~\cite{malleryRegularizedInverseHolographic2019}.

In contrast, optical diffraction tomography (ODT) \cite{Wolf1969, Devaney1981, Lauer2002, Charriere2006, Choi2007}, infers the 3D RI of a sample by combining multiple views obtained by projecting coherent light through the sample at different angles. At a high level, ODT is closely related to synthetic aperture microscopy~\cite{Alexandrov2006} and Fourier ptychography~\cite{Zheng2013}. ODT distinguishes itself from other ``volumetric" QPI approachs in its flexibility to probe arbitrary refractive indices and non-sparse samples and obtain true volumetric images. Therefore, ODT stands out as the method of choice for imaging dense samples in complex environments, as evidenced by a number of recent biological applications including measuring cell dry mass~\cite{lee_quantitative_2013}, flow cytometry~\cite{Ge2022}, traction force microscopy~\cite{Lee2023}, and 3D histopathology~\cite{hugonnet_multiscale_2021}. 

Due to the need to acquire $\gtrsim \num{100}$ views to perform high-quality tomographic reconstruction, ODT imaging is typically two orders of magnitude slower than single-shot techniques and achieving high-quality RI reconstructions at kilohertz volumetric imaging speeds is an open challenge. The primary hurdle to improve ODT acquisition speed is the rate of pattern generation for individual ODT views. Commonly, the angle that the illumination traverses the sample is controlled by galvanometric mirrors~\cite{Dong2020} or spatial light modulators (SLM's) \cite{Lee1979, Shin2015, Chamgoulov2004, Kuang2015, bianchiOpticalDiffractionTomography2022}. The fastest and most common SLM choice is the digital micromirror device (DMD), a binary device often run in a time-multiplexed gray-scale mode that limits the pattern refresh rate to \qty{1}{\kilo \hertz} \cite{Lee2017, Shin2018}. DMD's are capable of pattern changes at \qty{\geq 10}{\kilo \hertz} in binary mode, but this introduces stray diffraction orders that must be filtered out using static \cite{Shin2015} or dynamic \cite{Jin2018, Ge2022} masks. Even working at the full DMD display rate, achieving high-quality ODT using \num{>100} patterns limits the volumetric frame rate to \qty{<100}{\hertz}, still short of the volumetric kilohertz target. 

\begin{table}[ht]
\centering
\begin{tabular}{|l|l|l|l|l|l|}
\hline
Method & speed & limit & contrast & true 3D & ref\\
\hline
digital holographic microscopy & \qty{5}{\mega \hertz} & camera & RI & no & \cite{mazumdar_megahertz-rate_2020}\\
\hline
iSCAT & \qty{20}{\kilo \hertz} & camera & RI & no & \cite{brooks_point_2024}\\
\hline
SLIM & \qty{1}{\kilo \hertz} & camera/multiplexing & F & yes & \cite{wang_kilohertz_2024}\\
\hline
eventLFM & \qty{1}{\kilo \hertz} & sample/accumulation time & F & yes & \cite{Guo2024}\\
\hline
ODT & \qty{0.1}{\kilo \hertz} & DMD/camera & RI & yes & see text\\
\hline
FS-ODT & \qty{1}{\kilo \hertz} & DMD/multiplexing & RI & yes & this work\\
\hline
\end{tabular}
\caption{
\textbf{Comparison of \unit{\kilo \hertz}-scale volumetric imaging methods}.
RI and F stand for refractive index and fluorescence respectively. The quantitative phase imaging techniques are discussed in the text.                       SLIM and eventLFM are light field microscopy techniques.
\label{table:khz_methods}
}
\end{table}

To address the challenge of realizing high-resolution, true-3D, kilohertz-rate, and label-free volumetric imaging, we developed an alternative strategy to rationally designing and projecting ODT illumination patterns, Fourier synthesis optical diffraction tomography (FS-ODT). FS-ODT builds on our previous DMD pattern generation tools~\cite{Brown2021} to enable multiplexing many ODT views into a single image. We refer to this strategy as Fourier synthesis because we construct the complex illuminations patterns required for multiplexing by combining simple atomic patterns in the Fourier plane. Because FS-ODT places the DMD in a conjugate Fourier plane to the objective and achieves position and angle control over the patterns using a spatial carrier frequency, it is possible to increase the information content in a single image orders of magnitude further than multiplexing Lee holograms~\cite{Ge2022}. While  multiplexed intensity diffraction tomography (IDT) also synthesizes complex patterns using multiple LED point sources in the far-field~\cite{Tian2014, zhuHighfidelityIntensityDiffraction2022}, the use of a fast pattern modulator enables additional control over the beam phase and position and allows us to achieve two orders of magnitude faster volume acquisition.

Increasing the information content per image by multiplexing introduce a more ill-posed reconstruction problem. To address the ill-posed problem of multiplexed patterns, we designed an iterative reconstruction approach based on multi-slice beam propagation and accelerated proximal gradient descent. We additionally created a ``demultiplexing" approach to generate a high-quality initial 3D RI distribution and created new approaches to avoid non-optimal solutions due to phase wrapping.

We demonstrate the capabilities of FS-ODT for a variety of samples and scenarios, illustrating its unique combination of high-quality RI reconstruction and high-speed volumetric imaging capability. We first profile the reconstruction quality versus the degree of angle multiplexing by imaging a sample of known composition, a \qty{10}{\micro \meter} polystyrene microsphere (PMS) and a biological sample of known morphology, \textit{Tetrahymena}. Next, we consider dynamic samples, including diffusing PMS and bacteria, and demonstrate FS-ODT tracking of colloidal particles and extraction of their hydrodynamic properties. Finally, we combine position and angle multiplexing at the fastest FS-ODT imaging rate possible with our hardware to measure diffusing microspheres at kilohertz volumetric frame rates. Taken together, the results presented here demonstrate that FS-ODT is a powerful new approach for the dynamics of cells and colloids in complex environments.

\begin{figure}[htb!]
	\centering
	\includegraphics[width=\linewidth]{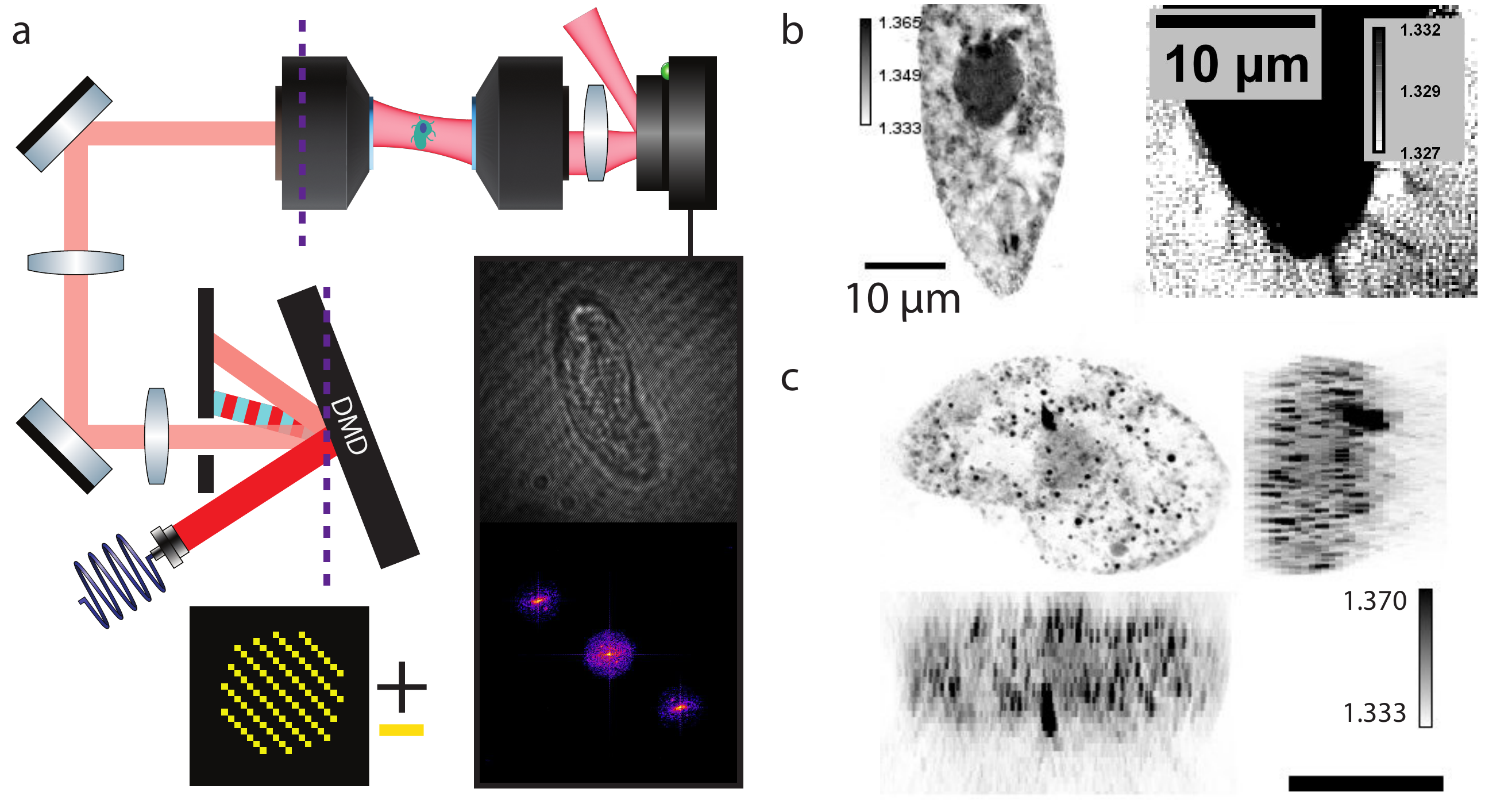}
	\caption{
    \textbf{Fourier synthesis optical diffraction tomography}.
    \textbf{a}. In FS-ODT, a spot-like pattern with a superimposed spatial carrier frequency is displayed on the DMD (lower left), which is conjugate to the back-focal plane of the objectives (purple line). The light which diffracts from the $-$ mirrors (red) at the carrier frequency is transmitted, while light which diffracts from the $+$ mirrors (blue) is blocked. The position and carrier frequency of the spot pattern control the angle and position of the ODT illumination respectively.
    \textbf{b}. Fixed \emph{Tetrahymena} imaged with FS-ODT shown in a single axial plane. Internal cell structures are resolved (left), and \qty{\sim 200}{\nano \meter} cilia are at the edge of our detection ability (right).
    \textbf{c}. MIP's of the \emph{Tetrahymena} RI in the $xy$- (upper left), $xz$- (lower left), and $zy$-planes (right). Internal structure of the cell can be resolved, including the nucleus and the oral apparatus. The unusual morphology suggests this cell may be in the process of dividing. Scale bar \qty{20}{\micro \meter}.    
    \label{fig:schematic}
    }
\end{figure}

\section*{Results}

\subsection*{Fourier synthesis of ODT patterns}

\begin{figure}[htb!]
	\centering
	\includegraphics[width=.8\linewidth]{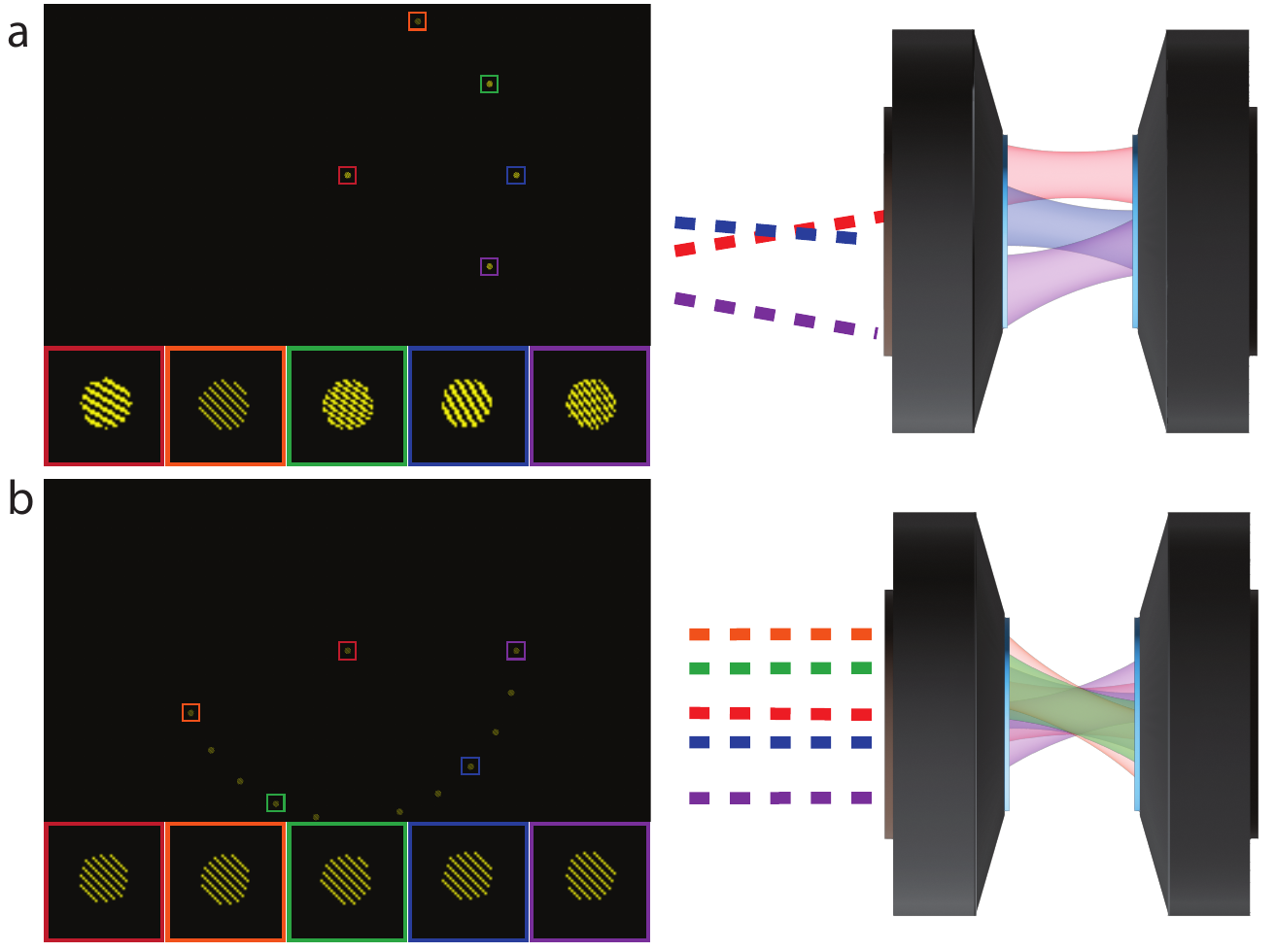}
	\caption{
    \textbf{Multiplexed illumination with FS-ODT}.
    \textbf{a}. Exemplary position multiplexing pattern. Five spot patterns are displayed across the DMD face (top left) with different carrier frequencies (inset, bottom left). When relayed to the objective back focal plane (right), the resulting beams have different incidence angles, and the lens transforms these angles to different positions in the focal plane. Only three beams are shown for clarity.
    \textbf{b}. Exemplary angle multiplexing pattern. 12 spot patterns (top left) are displayed at different spatial positions but with the same carrier frequency (inset, bottom left). When relayed to the objective back focal plane (right), the resulting beams have different incidence positions, and the lens transforms these positions to different angles in the focal plane. Only five beams are shown for clarity.
    \label{fig:pattern_synthesis}
    }
\end{figure}

In FS-ODT, we generate patterns using a DMD conjugate to the objective back focal (or Fourier) plane, as shown in Fig.~\ref{fig:schematic}a. Plane waves are encoded by circular ``spot'' patterns on the DMD. Varying the beam position in the sample corresponds to a adding a phase ramp in the Fourier plane, and so to make the beam position tuneable we align the system so that beams which diffract from a given spatial carrier frequency on the DMD are centered in the optical system. Beam angle through the sample is encoded by spot position, so multiplexing tens to hundreds of beams at different angles is possible by displaying multiple spatially separated spot patterns on the DMD. Further, by working in the Fourier plane we avoid the need to filter stray diffraction orders as these are mitigated by Fourier broadening due to the small spot sizes. We have carefully modelled this approach by extending the DMD simulation tools we developed previously for multicolor structured illumination microscopy~\cite{Brown2021} (section~\odtDetails).

To generate a single plane wave at spatial frequency $\vec{f} = (f_x, f_y)$ displaced from the center of the field of view by $\delta \vec{r}$, we generate a spot pattern at location $\vec{r}_p$ and spatial frequency $\vec{f}_p$ on the DMD. The object space parameters are determined by
\begin{align}
\delta \vec{r} &= \frac{f_l \lambda}{M} \left( \vec{f}_p - \vec{f}_c \right)\\
\vec{f} &= \frac{M}{f_l \lambda} \left( \vec{r}_p - \vec{r}_c \right)
\end{align}
where $\vec{f}_c$ is the spatial carrier frequency, $\vec{r}_c$ is the position on the DMD aligned with the optical axis, $f_l$ is the focal length of the objective lens, $M$ is the magnification between the DMD and the lens back focal plane, and $\lambda$ is the wavelength of light. This approach is power inefficient as the incident beam illuminates the full DMD chip, but spot patterns of typical radius \num{10} mirrors diffract only a small portion of the light. Nevertheless, since ODT detects transmitted light, there is sufficient signal to image with \qty{< 100}{\micro \second} exposure times (section~\dmdEff). 

To validate that the proposed FS-ODT optical design is capable of high-quality 3D RI reconstruction, we first performed non-multiplexed ODT on a complex 3D sample, fixed \emph{Tetrahymena} cells (Fig.~\ref{fig:schematic}b,c). After RI reconstruction, we resolve the internal structure of the cell, including the nucleus, the oral apparatus, and the cilia. The cilia are expected to be \qty{0.2}{\micro \meter} in diameter, which is significantly smaller than the Abbe limit for our detection optics, $2 \na / \lambda \sim \qty{0.39}{\micro \meter}$. As such, the measured refractive index contrast is small due to averaging of the structure and background over the resolution voxel size. Reconstruction parameters are provided in table~\ref{table:recon_params}.

We infer the RI distribution using custom GPU-accelerated Python code which implements accelerated proximal gradient descent using the fast iterative shrinkage-thresholding algorithm (FISTA)~\cite{Beck2009} (sections~\ref{section:code} and \reconstruction). We include the physics of light propagation through the RI using either the beam-propagation model (BPM) or the split-step non-paraxial (SSNP) forward models~\cite{Lim2019, zhuHighfidelityIntensityDiffraction2022}. FISTA and related proximal approaches are particularly powerful for solving ill-posed inverse image reconstruction problems because they provide a framework for applying regularization. Regularization stabilizes the reconstruction and favors physical solutions from all possible RI distributions consistent with the data. We use total variation regularization (TV), which promotes smoothness, and $\ell_1$ regularization, which promotes sparsity. Additionally, we impose physical constraints, typically that the imaginary part of the RI is strictly absorptive and the real part of the RI is greater than the background index.

Next, we apply the FS-ODT pattern generation strategy to generate multiplexed illumination patterns. Since FS-ODT uses spatially separated spot patterns to generate plane waves at different angles, we generate multiplexed patterns by displaying spot patterns at different positions on the DMD. We synthesize composite patterns from a set of beam frequency and position shift pairs $\left \{\left(\vec{f}_1, \delta \vec{r}_1 \right), ... \left(\vec{f}_N, \delta \vec{r}_N \right)  \right \}$. Each pattern generates $N$ beams passing through the sample plane in different positions at different angles (i.e. spatial frequencies). Unlike in multiplexed IDT approaches, these beams are mutually coherent, and so there combination generates a complex illumination pattern at the sample. Our approach enables two forms of multiplexing: (i) position multiplexing to extend the system field of view and (ii) angle multiplexing, which illuminates a single sample region with overlapping beams at different angles. We demonstrate these two strategies in Fig.~\ref{fig:pattern_synthesis}. A unique benefit of FS-ODT is that these two strategies can be combined to enable angle multiplexing over a large field of view. 

\subsection*{Low-resolution Rytov demultiplexing}

\begin{figure} [ht]
\begin{center}
\begin{tabular}{c}
\includegraphics[width=.6\textwidth]{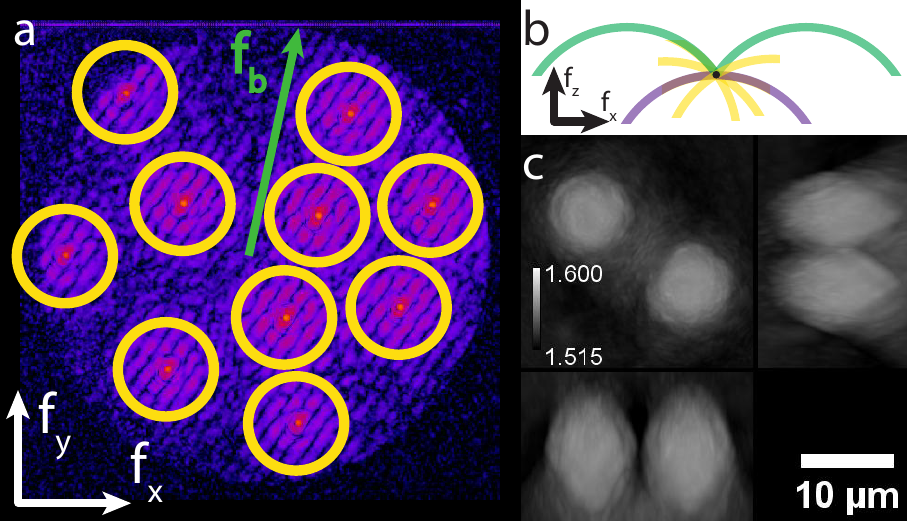}
\end{tabular}
\end{center}
\caption 
{ 
\textbf{Low-resolution Rytov demultiplexing}.
\textbf{a}. In the Fourier transform of each off-axis hologram, we select regions around each plane wave frequency where the scattered light is dominated by that individual plane wave (yellow circle). The radius of these regions is significantly smaller than the bandpass frequency of the microscope, $f_b$, so these regions contain low-resolution information about the RI object.
\textbf{b}. The information from each low-resolution region is mapped to the correct position (yellow arcs) in the 3D Fourier space representation of the scattering potential. These regions are centered about zero frequency (black dot). For reference we show arcs corresponding to a straight-on incident beam (purple) and beams at the band pass frequency (green).
\textbf{c}. We obtain an initial guess for the RI after inverse Fourier transforming the scattering potential. For the \qty{10}{\micro \meter} diameter PMS shown here, the low-resolution demultiplexed guess in the Rytov approximation provides an excellent starting point for further optimization.
\label{fig:low_res_rytov} 
}
\end{figure} 

Iterative ODT reconstruction, as described above, relies on a high-quality initial guess for the RI distribution. Since the problem is not globally convex, the iteration scheme may not converge if the guess is too far from the true RI distribution. The initial guess is most commonly generated from linear weak scattering approximations, such as the Born and Rytov approximations, that can be efficiently computed from the electric field data. For plane wave illumination, these approximations relate the 2D Fourier transform of the image electric field to a spherical cap in the 3D Fourier transform of the scattering potential (section~\linscatt).

Multiplexing many patterns results in a more ill-posed inverse problem than single-beam ODT due to the additional need to unmix scattering contributions from different incident plane waves.  When the illumination pattern contains multiple plane waves, it is not obvious which spherical cap each image Fourier frequency should be assigned. This ambiguity can be removed experimentally using phase shifting, but this requires additional images and negates the speed advantage of multiplexing.

To improve the convergence of our reconstruction algorithm, we develop an approach to initialize the RI with a high-quality guess in the presence of multiplexing, which we refer to as low-resolution Rytov demultiplexing. In this scheme, we observe that in many common experimental situations, information from the scattering of $i$th plane wave dominates the hologram in the region around the plane wave carrier frequency. We expect this situation to prevail if scattering is weak enough and, for example, the sample RI is greater than the background and its power spectrum decays with increasing spatial frequency. For each plane wave, we select the region in Fourier space which is closer to that carrier frequency than to any other. We take the resulting $N$ lower-resolution demultiplexed fields and compute the corresponding Rytov phases (eq.~\rytovPhase). Finally, we generate an initial RI distribution by using the linear scattering model to map this information to the scattering potential at the correct position in Fourier space.  An example of this process is shown in Fig.~\ref{fig:low_res_rytov}a, b for $10\times$ multiplexing for a sample consisting of \qty{10}{\micro \meter} diameter PMS. The resulting Rytov demultiplexed RI (Fig.~\ref{fig:low_res_rytov}c) captures the features of the sample and is a good initial guess for the solver.

While we find Rytov demultiplexing produces a better initial guess than e.g. beginning with a constant RI, it eventually fails because as the degree of multiplexing increases, the resolution of the demultiplexed images decreases. For small objects and high degrees of multiplexing, the demultiplexed images do not contain sufficient detail to accurately represent the RI.  This problem is particularly acute for thick objects that produce phase-wrapped electric fields, because the phase wraps must be detected by comparing phase changes across the object. When rapid phase variations are not captured by the low-resolution demultiplexed images, phase unwrapping breaks down and phase errors in the initial guess can lead to rapid deterioration of the RI reconstruction.

In some cases, the sensitivity of the optimization problem to correct phase-unwrapping can be mitigated by using a loss function which includes a phase-sensitive and a phase-insensitive term (eq.~\lossFnMix). This illustrates that, while the phase information obtained by ODT produces a less ill-poised reconstruction problem it simultaneously produces a more complex landscape for the loss function which poses significant challenges for convex optimization approaches.

\subsection*{FS-ODT validation}
\begin{figure}[htb!]
	\centering
	\includegraphics[width=\linewidth]{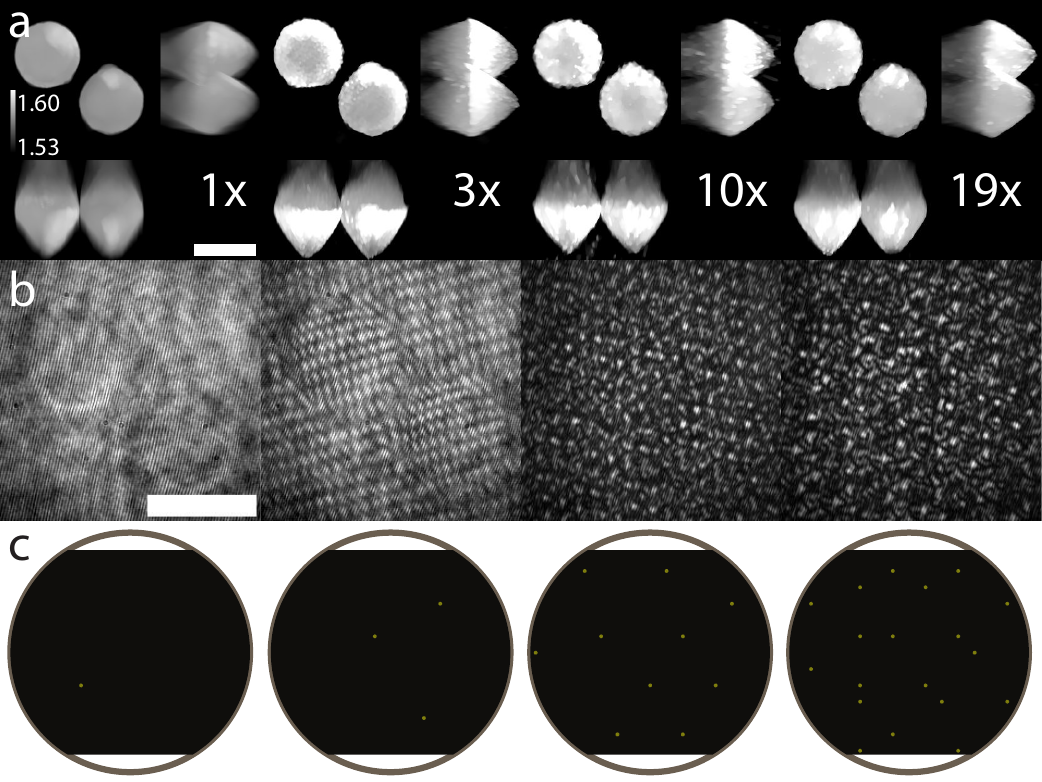}
	\caption{
    \textbf{\qty{10}{\micro \meter} microsphere refractive index reconstruction versus angle multiplexing}.
    \textbf{a}. RI reconstruction using the SSNP and an initial guess from the low-resolution demultiplexed Rytov approach described in the text. MIP's through the microsphere are shown in the $xy$-plane (upper left), $yz$-plane (upper right), and $xz$-plane (lower left). Results for angle multiplexing factors of \num{1}, \num{3}, \num{10}, and \num{19} are shown (left to right). Scale bar \qty{10}{\micro \meter}
    \textbf{b}. Exemplary raw hologram images which are used to recover the bead RI. Scale bar \qty{10}{\micro \meter}. For higher degrees of multiplexing, these become more complex due to the mutual interference of many plane waves.
    \textbf{c}. Exemplary DMD patterns used to generate the multiplexed illumination patterns in B. Only the portion of the DMD within the detection objective pupil radius (grey circle) is shown. The DMD is slightly smaller (\qty{\sim 84}{\percent}) than the pupil along the vertical direction.
    \label{fig:10um_bead}
    }
\end{figure}

As an initial validation of our approach, we collect FS-ODT images of a sample of known structure and RI, a \qty{10}{\micro \meter} diameter PMS suspended in immersion oil, using a range of different multiplexing conditions. We consider multiplexing by factor of \num{1}, \num{3}, \num{10}, \num{19} using a fixed set of $N = \num{147}$ plane wave directions. Where the total number of angles is not divisible by the multiplexing factor, we include some plane waves twice to simplify the reconstruction.

While FS-ODT can generate arbitrary multiplexed patterns, reconstruction quality is improved using multiplexed patterns that are easier to unmix by ensuring the beam frequencies in each image are as far apart as possible. This maximizes the resolution achieved in the Rytov demultiplexing initialization. We design our patterns using an iterative algorithm to ensure this is the case. We initialize each $M$-fold multiplexed pattern with $M$ beam angles by iteratively selecting the beam angle that maximizes the distance from those already chosen. Next, we iterate over all choices of two patterns and randomly swap beam angles. We define a loss function, the average distance between beam angles on the DMD up to a certain maximum value. If swapping the angles increases the loss function, we keep the swap. We obtain high-quality multiplexed patterns after performing \num{5} iterations of \num{300} swaps.

Due to the thickness and large RI contrast between the PMS and the immersion oil, $n_\text{PMS} \sim 1.57$ and $n_o \sim 1.515$, we use the SSNP for our reconstruction~\cite{Lim2019} to both allow for multiple scattering and accurately model the propagation of the complex illumination beam. The BPM with an obliquity factor correction achieves similar performance with reduced memory and computational requirements for single plane wave illumination~\cite{zhuHighfidelityIntensityDiffraction2022}, but in our case the multiple incident plane waves do not share a common obliquity factor, so the SSNP is more appropriate. 

We find that FS-ODT achieves high-quality 3D RI reconstructions of PMS for $1\times$ (no multiplexing), $3\times$, $10\times$, and $19\times$ multiplexing (fig.~\ref{fig:10um_bead}). With no multiplexing, we recover a nearly spherical object of the correct diameter with $n \approx 1.57$, as expected because this problem is the least ill-posed and has the largest effective SNR per beam. Since all raw images are acquired with similar peak intensity, the effective SNR per beam is reduced as the degree of multiplexing increases. As expected, due to the missing cone problem, the sphere appears somewhat elongated in the axial direction. The $3\times$, $10 \times$, and $19 \times$ multiplexed data also produces high-quality reconstructions, but the RI is not distributed as uniformly along the optical axis. This reconstruction artifact is reminiscent of missing-cone effects, and could be a result of effective loss of SNR for individual plane waves in the highly-multiplexed beam.

We numerically explore and further validate our multiplexed FS-ODT approach by reconstructing synthetic images generated using custom GPU accelerated Mie scattering code (section~\mieRecon). We find that multiplexed patterns significantly enhance the RI reconstruction compared with non-multiplexed patterns when the number of raw tomograms is kept fixed.

\subsection*{Hindered diffusion and microswimmer motility}

\begin{figure}[htb!]
	\centering
	\includegraphics[width=\linewidth]{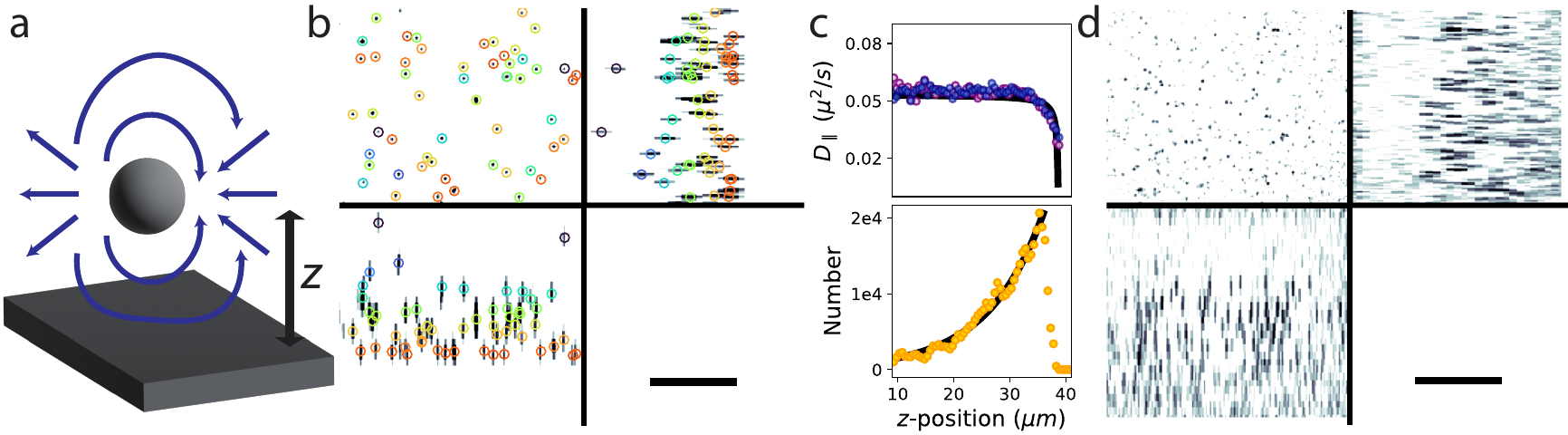}
	\caption{
    \textbf{Hindered diffusion of colloidal microspheres}.
    \textbf{a}. Colloidal particles interact with nearby boundaries when the boundary influences the fluid flow fields generated by the particles' motion.
    \textbf{b}. \qty{1}{\micro \meter} diameter PMS diffusing in water-glycerol mixture imaged with FS-ODT. MIP's of the RI are shown in three orthogonal planes. Circles indicate bead location determined by tracking algorithm. Color represents axial position from near the coverslip (red) to up in the sample (purple). Scale bar \qty{20}{\micro \meter}.
    \textbf{c}. Hydrodynamic interactions between microspheres and the coverslip induce a spatially varying diffusion coefficient (top). The microspheres are buoyant in this solution, leading to a spatially varying density profile (bottom).
    \textbf{d}. FS-ODT resolves individual microspheres in a solution with $\times 10$ greater PMS concentration compared with b. Scale bar \qty{20}{\micro \meter}.
    \label{fig:hydro_int}
    }
\end{figure}

\begin{figure}[htb!]
	\centering
	\includegraphics[width=.8\linewidth]{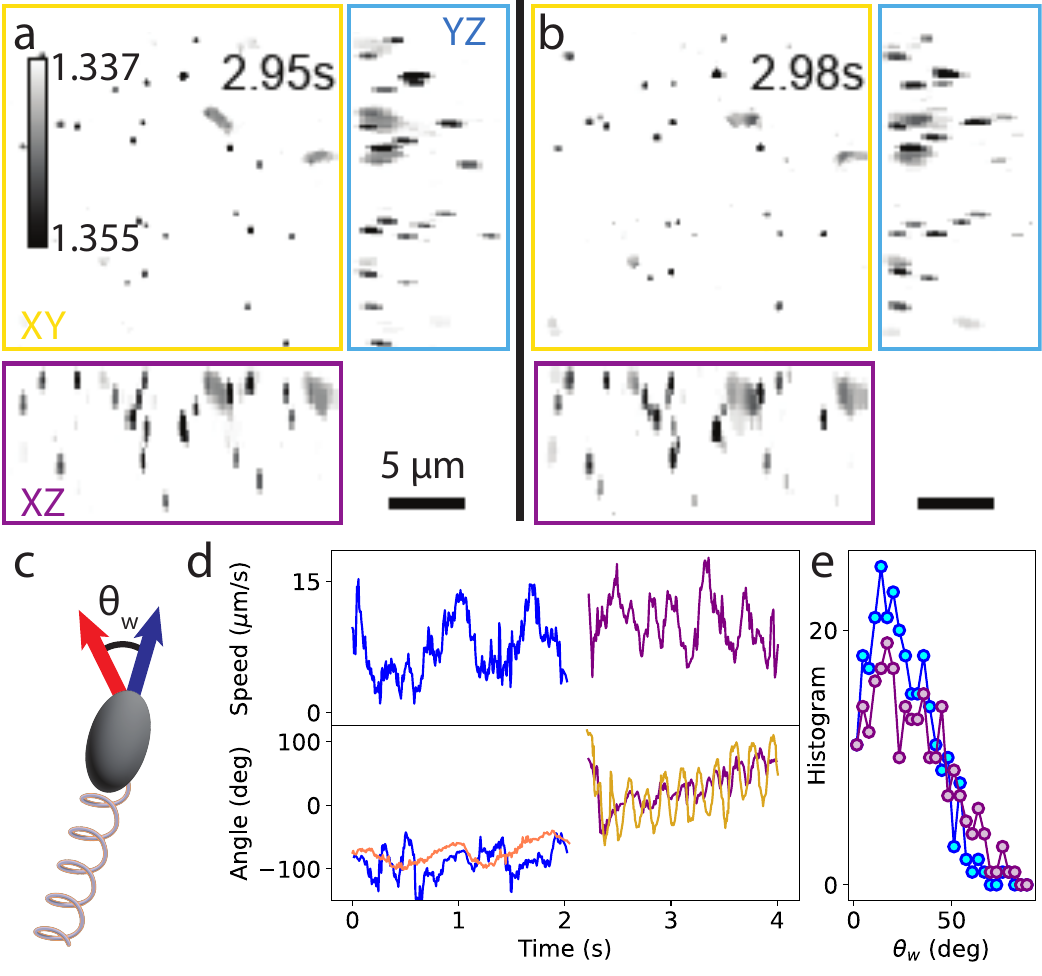}
	\caption{
    \textbf{\emph{E. coli} motility}.
    \textbf{a}. RI reconstruction of bacteria and tracer particles MIP along three orthogonal axes, $XY$ (upper left), $YZ$ (right), and $XZ$ (bottom). The bacteria and tracers can be distinguished based on their RI and morphology. Two bacteria (upper right) and \num{\sim 20} tracer particles are visible in the $XY$ projection.
    \textbf{b}. Bacteria and tracer particles \qty{30}{\milli \second} later. One bacteria has rotated, and the tracers have changed position due to diffusion and advection.
    \textbf{c}. During bacterial swimming we define the body axis (blue) and the average velocity (red). The wobble angle, $\theta_w$, is the average angle between the two.
    \textbf{d}. Speed (top) and 2D orientation of the bacteria axis (red and gold) and the bacteria velocity (blue and purple). Different traces represent the two swimming bacteria observed in the experiment.
    \textbf{e}. Histogram of measured wobble angles. Colors correspond with bacteria identified in d.
    \label{fig:ecoli}
    }
\end{figure}

Having established FS-ODT produces high quality RI reconstructions in static samples, we now apply our approach to study hindered diffusion, hydrodynamic interactions, and microswimmer motility. Phase imaging approaches offer significant advantages over fluorescence imaging in lower phototoxicity, faster frame rates, and axial position sensitivity. As such, phase methods are ideally suited to studying motion in 3D environments, such as the motility of colloids or cells in complex environments. One poorly understood question is how cells swim in viscoelastic environments created by the presence of inert colloids or polymers in solution~\cite{kamdarColloidalNatureComplex2022}. This topic is of broad interest because these fluids mimic the natural environment of bacterial species better than typical \emph{in vitro} experiments and may provide insight into behavior and evolutionary pathways which is currently lacking. Complex environments may also impact collective bacterial dynamics, including swarming and biofilm formation, which are interesting from a fundamental perspective of better understanding active matter as well as from a public health perspective~\cite{kostakioti_bacterial_2013}. 3D QPI approaches offer significant advantages over 2D-only or 2D with limited $z$-range tracking, which only measure particles at or near the microscope's focal plane, particularly in dense environments like bacterial suspensions in complex fluids. As a first step towards addressing these questions, we demonstrate that we can image and distinguish diffusing microspheres and swimming \emph{E. coli} cells in 3D using FS-ODT.

Imaging dense samples introduces new reconstruction challenges, as our algorithms require knowledge of the excitation electric field with no sample present to distinguish electric field patterns generated by the complex illumination pattern from those due to interaction with RI objects. One standard solution is to acquire a background image taken at a spatial position where no RI objects are present. However, for dense diffusing objects, there are no sample-free regions. Instead, we rely on the time-average image, assuming that over the long term, the RI inhomogeneities average out. For relatively sparse samples, this is a good approximation. For denser samples, it may be necessary to account for an average background RI different from the fluid. Other approaches, including joint inference of the incident electric field and the sample RI, are possible~\cite{Thibault2009}.

We applied FS-ODT to study hindered diffusion of \qty{1}{\micro \meter} beads in a water glycerol mixture and quantified the bead dynamics by determining the diffusion coefficient as a function of distance to the coverslip using 3D tracking (Fig.~\ref{fig:hydro_int}, Videos~\videoTrackingBeads~and \videoTrackingBeadsDense). The diffusion coefficient is sensitive to hydrodynamic forces on the beads and serves as an indicator of hydrodynamic interactions. Colloidal particles experience hydrodynamic interactions with boundaries because they generate fluid flow fields as they move, which must satisfy the Stokes equation with appropriate boundary conditions. When a sphere is within a few radii, $R$, of a boundary, the no-slip boundary condition at the wall modifies the flow fields (Fig.~\ref{fig:hydro_int}a) which affects the mobility of the sphere and, therefore, the translational and rotational diffusion coefficients. This effect, known as hindered diffusion, is perturbative in the ratio of the sphere's radius to the distance from the wall $R/h$. To leading order, the diffusion coefficient parallel to the wall is~\cite{Brenner1983}
\begin{align}
    D_\parallel &= D_o \left[1 - \frac{9}{16} \frac{R}{h} + \mathcal{O} \left( \frac{R^2}{h^2} \right) \right],
\end{align}
where $D_o$ is the diffusion coefficient in a unbounded fluid. Our measured diffusion coefficients versus height above the coverslip (Fig.~\ref{fig:hydro_int}b) match well this functional form. There are small deviations from the expected curvature, which we attribute to the fact that the microspheres sample different heights during our measurement, leading to some broadening of the measured curve. 

Hindered diffusion has previously been studied using evanescent wave dynamical light scattering~\cite{holmqvist_colloidal_2007} and optical tweezer experiments combined with in-line holography~\cite{sharma_high_precision_2010}. FS-ODT provides an alternative approach which can resolve hydrodynamic forces over tens of \unit{\micro \meter} axially without the need to perturb the sample, scan tweezer positions, or combine multiple imaging techniques. ODT's unique ability to image across a 3D field of view and in multiple scattering environments enables simultaneously tracking of many particles in dense colloidal suspensions, which is extremely challenging using these other approaches.

Next, we studied \emph{E. coli} motility by imaging bacteria in solutions seeded with \qty{0.5}{\micro \meter} PMS tracer particles at volumetric rates of \qty{143}{\hertz} over \qty{\sim 7}{\second} (Fig.~\ref{fig:ecoli}, Videos~\videoEcoli~and \videoEcoliThreeD). We found that the bacteria and tracer particles can be distinguished based on RI and morphology. We observe diffusion of the tracer particles and directed swimming of \num{2} \emph{E. coli} cells in this dataset. To quantify the swimming behavior of the \emph{E. coli}, we trained a classifier to segment the cell bodies and then tracked their position and orientation (Fig.~\ref{fig:ecoli}D). From these 4D data, we computed the wobble angle, which is the average angle between the cell's average velocity and body axis (Fig.~\ref{fig:ecoli}c). Wobble angles have previously been used to characterize \emph{E. coli} swimming in complex fluids~\cite{kamdarColloidalNatureComplex2022}. This experiment demonstrates FS-ODT can effectively identify bacteria swimming in 3D and be used to compute wobble angles, which we anticipate will enable larger-scale studies.

\subsection*{Diffusing microspheres imaged at kilohertz volumetric rates}
\begin{figure}[htb!]
	\centering
	\includegraphics[width=\linewidth]{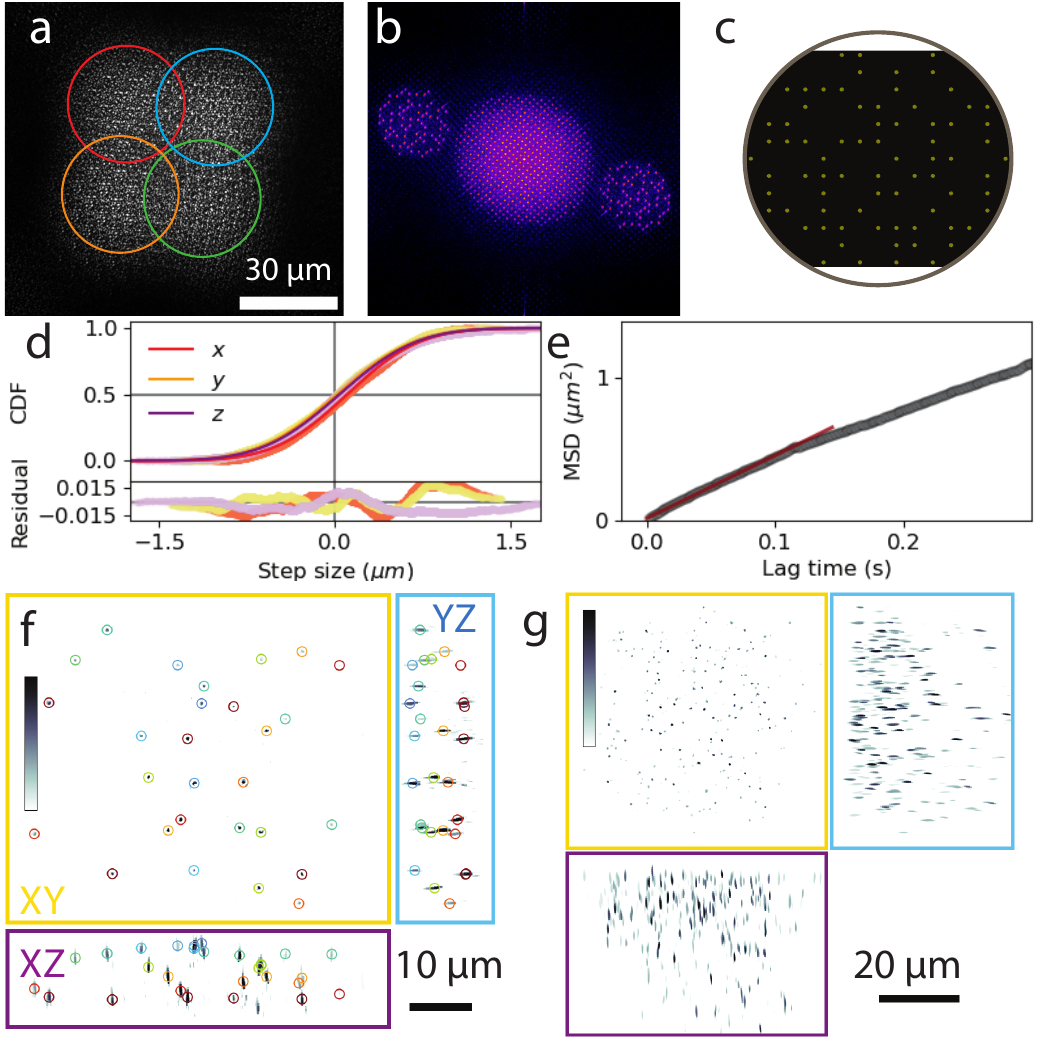}
	\caption{
    \textbf{Diffusing microspheres at kilohertz volumetric rates with an extended field-of-view}.
    \textbf{a}. Intensity profile of the beam after interacting with microspheres using $19 \times$ angle multiplexing and $4 \times$ position multiplexing (circles). Position multiplexing expands the usable field of view by $\sim 2 \times$
    \textbf{b}. Fourier transform of the intensity profile in a showing \num{76} distinct beams in Fourier space.
    \textbf{c}. DMD pattern used to generate the illumination shown in a and b
    \textbf{d}. Step-size cumulative distribution function (points), Gaussian model fits (lines), and residuals (bottom) for steps separated by \num{200} frames along the $x$ (red-orange), $y$ (yellow), and $z$ (purple) directions.
    \textbf{e}. Ensemble-average mean-square displacement (MSD) versus lag-time (grey) and a linear fit (red) up to a lag-time of \qty{0.144}{\second}.
    \textbf{f}. RI reconstruction at a single time point for a \qty{\sim 10}{\micro \meter} tall chamber. RI values are shown between \num{1.352} and \num{1.413}. PMS are localized (circles) and tracked.
    \textbf{g}. RI reconstruction at a single time point for a denser sample in a \qty{120}{\micro \meter} tall chamber. RI scale the same as f.
    \label{fig:diffusing_beads}
    }
\end{figure}

We next demonstrate that FS-ODT can reliably image diffusing colloidal particles at kilohertz volumetric rates. Specifically, we image \qty{1}{\micro \meter} PMS diffusing in water over a period of \qty{0.969}{\second} at a volumetric rate of \qty{1.032}{\kilo \hertz} with a field-of-view of \qtyproduct{20.5 x 96 x 104 x}{\micro \meter} (Videos~\videoKhzBeads~and \videoKhzBeadsDenser). At each time point we collect \num{8} images using $19 \times$ angle multiplexed patterns. Additionally, we expand the field-of-view of the imaging system by expanding the illumination pattern $2 \times$ in each direction by including spots with \num{4} different carrier frequencies. This combined angle and position multiplexing uses a total of \num{608} plane wave patterns.  We show an exemplary angle- and position-multiplexed illumination pattern in Fig.~\ref{fig:diffusing_beads}a--c.

We performed RI reconstruction at each time point (Fig.~\ref{fig:diffusing_beads}f) and tracked the diffusing PMS to extract their diffusion coefficients. As for other time-resolved experiments, we construct a reference electric field by time-averaging the electric fields obtained from off-axis holography. We quantify the diffusive motion by calculating cumulative distribution functions (CDF's) of the step-size distribution along each axis (Fig.~\ref{fig:diffusing_beads}d) and the mean-square displacement (MSD) of the beads versus lag-time (Fig.~\ref{fig:diffusing_beads}e). For short lag-times, the CDF's can be strongly affected by localization error. Therefore we examine the CDF's for a lag-time of \qty{0.194}{\second} (200 frames) and find that the CDF's closely match the expected Gaussian form. We extract the ensemble-average diffusion coefficient by performing a linear fit to the MSD and determine $D = \qty{0.68}{\micro \meter^2 \per \second}$.

We display volumetric data in Fig.~\ref{fig:diffusing_beads} using maximum intensity projections (MIP's) along three orthogonal axes. We can easily resolve the diffusing PMS, and we find that they are distributed over an axial range of about \qty{10}{\micro \meter}, which is the approximate height of the sample chamber (Fig.~\ref{fig:diffusing_beads}f). We also imaged PMS diffusing in a \qty{120}{\micro \meter} thick chamber, focusing on the region within \qty{40}{\micro \meter} of the coverslip (Fig.~\ref{fig:diffusing_beads}g). These images provide an ideal starting point for particle tracking, particle imaging velocimetry, or other approaches to either explore the physics of colloidal suspensions or use these colloids as tracer particles to report on more complex biological systems.

\section*{Discussion}
Optical diffraction tomography is a powerful label-free 3D imaging approach with broad applications in biomedical imaging, biophysics, and soft matter physics. However, most efforts focus on improving ODT for applications specific to biomedical imaging. Biophysics and soft matter systems have very different imaging requirements, motivating our development of multiplexed FS-ODT pattern generation and reconstruction approaches.

We demonstrated that FS-ODT provides high-quality RI reconstructions of a variety of samples, including cells, \qty{10}{\micro \meter} microspheres, diffusing colloids, and swimming bacteria. We present multiplexed illumination strategies to increase the number of illumination angles in a single image and expand the system's field of view. Combined, both multiplexing strategies enable \unit{\kilo \hertz}-scale ODT over extended lateral fields view. 

While the \unit{\kilo \hertz}-scale volumetric imaging is fast enough for many applications, FS-ODT imaging could be further accelerated by adopting alternative pattern generation hardware. For example, using faster DMD models that have pattern update rates up to \qty{30}{\kilo \hertz} could immediately accelerate the volumetric imaging rate by a factor of \num{3}. Alternatively, acousto-optic deflectors or electro-optic modulators could accelerate pattern generation by \numrange{1}{2} orders of magnitude, at the cost of complicating the off-axis holography reference generation. These fast speeds might be necessary for studying swimming in fast ciliates, which can move at speeds of up to \qty{4000}{\micro \meter / \second}~\cite{lisicki_swimming_2019} or studying the diffusion of single proteins~\cite{brooks_point_2024}.

While FS-ODT already produces high-quality reconstructions in most circumstances, further improvements are possible by applying FISTA or a deep learning approach to unmix the off-axis holograms, denoise raw data, suppress coherent speckle, or perform phase unwrapping. Supervised learning approaches may improve the speed or quality of multiplexed reconstructions ~\cite{matlockMultiplescatteringSimulatortrainedNeural2023, Ge2022}.

Alternatively, emerging deep-learning regularization approaches could be incorporated to improve FS-ODT reconstructions further. For example, DNN denoisers can be incorporated as priors in our existing FISTA framework using the plug-and-play prior or regularization by denoising approaches~\cite{wu_simba_2020}. Alternatively, recent self-supervised approaches have achieved impressive results while preserving a physics-based forward model by using DNN's to parameterize the RI distribution, either as a pixelated image as in deep image prior approaches~\cite{Bostan2020} or as a function of coordinates, as in neural field based approaches~\cite{liu_recovery_2022}. To account for motion blur, it may be possible to adapt neural field approaches to create a space-time model and infer dynamics during hologram acquisition~\cite{cao_dynamic_2022}.

We anticipate that FS-ODT will unlock challenging imaging regimes that have been difficult or impossible to explore. These include probing the \qty{\sim 100}{\hertz} mechanical motion and hydrodynamics of cilia during swimming of protists such as \emph{Tetrahymena} or exploring the behavior of microswimmers and active colloids in dense or complex 3D environments. It also extends the frontier of QPI to complex object tracking in dense environments and the exploration of highly dynamic systems.

\section*{Acknowledgements}
We thank Siqi Yang and Shwetadwip Chowdhury for their valuable discussions on ODT pattern generation, pattern multiplexing, and reconstruction. We thank Jessi Vlcek and Steve Press{\'e} for providing \emph{E. coli} samples for instrument validation. PTB and DPS acknowledge funding from Scialog, Research Corporation for Science Advancement, and Frederick Gardner Cottrell Foundation 28041 and grant number 2021- 236170 from the Chan Zuckerberg Initiative DAF, an advised fund of the Silicon Valley Community Foundation. Portions of this work were previously presented and published in conference proceedings~\cite{brown_fourier_2023, brown_spie_2024}.

\section*{Data Availability}
Instrument control and ODT reconstruction code are available on GitHub, and the version used in this work is archived on Zenodo~\cite{mcsim}. Raw image data and RI reconstructions for Figs.~\ref{fig:schematic} and \ref{fig:10um_bead}, MIP's for Figs.~\ref{fig:hydro_int}, \ref{fig:ecoli}, and \ref{fig:diffusing_beads}, segmentation masks for Fig.~\ref{fig:ecoli}, and tracking data from Figs.~\ref{fig:hydro_int}, \ref{fig:ecoli}, and \ref{fig:diffusing_beads} is available on Zenodo \url{https://doi.org/10.5281/zenodo.10883021}. Time-lapse data and RI reconstructions from Figs.~\ref{fig:hydro_int}, \ref{fig:ecoli}, and \ref{fig:diffusing_beads} are available from the authors upon reasonable request.

\section*{Conflict of Interest}
The authors declare no competing interests.

\section*{Author contributions}
PTB and DPS conceived the project and prepared the manuscript. PTB built the instrument, wrote the control and reconstruction code, and performed the experiments. DPS supervised all aspects of the project. NJ developed the \emph{E. coli} tracking code. RK provided assistance with sample preparation. LM and NW prepared the \emph{E. coli}. EM, AP, and DFG prepared the \emph{Tetrahymena}.

\section*{Materials and Methods \label{section:sample_prep}}
\begin{table}[ht]
\centering
\resizebox{\columnwidth}{!}{%
\begin{tabular}{|l|l|c|c|c|c|c|l|l|c|c|c|c|c|c|c|c|}
\hline
sample & figure & $M$ & $N_p$ & $N_t$ & rate & exp & model &  $n_o$ & $f$ & $\tau_\text{tv}$ & $\tau_{\ell_1}$ & max $n''$ & batch & $dz \times dxy$ & FOV ($z \times y \times x$) & $T_\text{recon}$\\
\hline 
\emph{Tetrahymena} & Fig.~\ref{fig:schematic}b & 331 & 331 & 1 & -- & \qty{1}{\milli \second} & Rytov & \num{1.333} & 1 & 0.1 & 0.01 & $\infty$ & 1 & \qtyproduct{0.869 x 0.202}{\micro \meter} & \qtyproduct{32.2 x 47.6 x 60.9}{\micro \meter} &\\
\hline
\emph{Tetrahymena} & Fig.~\ref{fig:schematic}c & 145 & 145 & 1 & -- & \qty{600}{\micro \second} & BPM & \num{1.333} & 1 & \num{0.03} & 0 & $\infty$ & 3 & \qtyproduct{0.87 x 0.2}{\micro \meter} & \qtyproduct{26.97 x 50.4 x 56}{\micro \meter} & \qty{3}{\minute}\\
\hline
\qty{10}{\micro \meter} PMS $1\times$ & Fig.~\ref{fig:10um_bead} & 147 & 147 & 1 & -- & \qty{400}{\micro \second}&  SSNP & \num{1.515} & 1 & 0.1 &  0.003 & 0 & 4 & \qtyproduct{0.1 x 0.1}{\micro \meter} & \qtyproduct{25.1 x 70 x 70}{\micro \meter} & \qty{20}{\minute}\\
\hline
\qty{10}{\micro \meter} PMS $3\times$ & Fig.~\ref{fig:10um_bead} & 49 & 147 & 1 & -- & \qty{300}{\micro \second} & SSNP & \num{1.515} & 1 & 0.1 & 0.003 & 0 & 4 & \qtyproduct{0.1 x 0.1}{\micro \meter} & \qtyproduct{25.1 x 70 x 70}{\micro \meter} & \qty{16}{\minute}\\
\hline
\qty{10}{\micro \meter} PMS $10\times$ & Fig.~\ref{fig:10um_bead} & 15 & 150 & 1 & -- & \qty{150}{\micro \second} & SSNP & \num{1.515} & 0.1 & 0.1 & 0 & 0 & 2 & \qtyproduct{0.1 x 0.1}{\micro \meter} & \qtyproduct{25.1 x 70 x 70}{\micro \meter} & \qty{32}{\minute}\\
\hline
\qty{10}{\micro \meter} PMS $19\times$ & Fig.~\ref{fig:10um_bead} & 8 & 152 & 1 & -- & \qty{75}{\micro \second} & SSNP & \num{1.515} & 0.01 & 0.1 & 0 & 0 & 2 & \qtyproduct{0.1 x 0.1}{\micro \meter} & \qtyproduct{25.1 x 70 x 70}{\micro \meter} & \qty{32}{\minute}\\
\hline
\qty{1}{\micro \meter} PMS & \begin{tabular}{l} Fig.~\ref{fig:hydro_int}b \\ Video~\videoTrackingBeads \end{tabular} & 11 & 11 & \num{10000} & \qty{6.05}{\hertz} & \qty{3}{\milli \second} & Rytov & \num{1.407} & 1 & 0 & 0 & $\infty$ & 11 & \qtyproduct{1.75 x 0.376}{\micro \meter} & \qtyproduct{\sim 54 x 67 x 79}{\micro \meter} & \qty{14}{\hour}\\
\hline
\qty{1}{\micro \meter} PMS & \begin{tabular}{l} Fig.~\ref{fig:hydro_int}d \\ Video~\videoTrackingBeadsDense \end{tabular} & 11 & 11 & \num{2000} & \qty{6.05}{\hertz} & \qty{3}{\milli \second} & Rytov & \num{1.407} & 1 & 0 & 0 & $\infty$ & 11 & \qtyproduct{1.75 x 0.376}{\micro \meter} & \qtyproduct{\sim 54 x 67 x 79}{\micro \meter} &\\
\hline
\begin{tabular}{l} \qty{0.5}{\micro \meter} PMS\\ \& \emph{E. coli} \end{tabular} & \begin{tabular}{l} Fig.~\ref{fig:ecoli}c,d \\ Video~\videoEcoli\\Video~\videoEcoliThreeD \end{tabular} & 11 & 11 & \num{1000} & \qty{143}{\hertz} & \qty{600}{\micro \second} & BPM & \num{1.333} & 1 & 0.03 & 0.01 & $\infty$ & 11 & \qtyproduct{0.435 x 0.196}{\micro \meter} & \qtyproduct{25.6 x 25.3 x 26.5}{\micro \meter} & \qty{49}{\minute}\\
\hline
\qty{1}{\micro \meter} PMS & \begin{tabular}{l} Fig.~\ref{fig:diffusing_beads}a-f\\ Video~\videoKhzBeads \end{tabular} & 8 & 608 & \num{1000} & \qty{1.032}{\kilo \hertz} & \qty{75}{\micro \second} & BPM & \num{1.333} & 1 & \num{0.01} & \num{0.0003} & 0 & 3 & \qtyproduct{0.5 x 0.1}{\micro \meter} & \qtyproduct{40.5 x 96 x 104}{\micro \meter} & \qty{48}{\hour}\\
\hline
\qty{1}{\micro \meter} PMS & \begin{tabular}{l} Fig.~\ref{fig:diffusing_beads}g\\ Video~\videoKhzBeadsDenser \end{tabular} & 8 & 608 & \num{1000} & \qty{1.032}{\kilo \hertz} & \qty{75}{\micro \second} & BPM & \num{1.333} & 1 & \num{1} & \num{0.01} & 0 & 3 & \qtyproduct{0.5 x 0.1}{\micro \meter} & \qtyproduct{70.5 x 96 x 104}{\micro \meter} & \qty{66}{\hour}\\
\hline
\qty{10}{\micro \meter} Mie $10 \times$ & Fig.~\mieRecon a & 15 & 150 & 1 & -- & -- & SSNP & 1.515 & 1 & 1 & 0 & 0 & 1 & \qtyproduct{0.1 x 0.1}{\micro \meter} & \qtyproduct{25.1 x 70 x 65}{\micro \meter} &\\
\hline
\qty{10}{\micro \meter} Mie $1 \times$ & Fig.~\mieRecon b & 15 & 15 & 1 & -- & -- & SSNP & 1.515 & 1 & 1 & 0 & 0 & 1 & \qtyproduct{0.1 x 0.1}{\micro \meter} & \qtyproduct{25.1 x 70 x 65}{\micro \meter} &\\
\hline
\end{tabular}
} 
\caption{\textbf{ODT reconstruction parameters}. Here we provide details about the ODT patterns and reconstruction parameters used in all data sets presented above. The pattern parameters are $M$, the number of images, $N_p$, the number of plane wave patterns, and $N_t$, the number of time lapse points collected. Rate is the volumetric image acquisition rate and exp is the exposure time for each off-axis hologram. The reconstruction parameters are the fraction of electric field loss function, $f$, the regularization parameters $\tau_\text{tv}$ and $\tau_{\ell_1}$,  the maximum allowed value for the imaginary part of the RI, max $n''$, the voxel size of the reconstruction grid, $dz \times dxy$, and the reconstruction field of view, FOV. $T_\text{recon}$ is the reconstruction time.
}
\label{table:recon_params}
\end{table}

\subsection{Computer control \label{section:code}}
The microscope is controlled using a custom computer package~\cite{mcsim}, and a GUI based on Napari-MicroManager, which relies on the MicroManager device drivers but replaces the Java GUI with Napari, \url{https://github.com/QI2lab/napari-micromanager}. The control computer is a Lenovo ThinkStation P620 running Microsoft Windows 10 Enterprise. This computer has an Ryzen Threadripper Pro 3945WX CPU (AMD) with \num{12} cores, \qty{128}{\giga \byte} RAM, and an GeForce RTX 3090 GPU (NVidia) with \qty{24}{\giga \byte} of memory. ODT patterns are pre-loaded on the DMD firmware, and the entire microscope acquisition is hardware triggered by a National Instruments DAQ (PCIe-6343). The Phantom camera includes \qty{72}{\giga \byte} on-board memory and the 10 Gigabit ethernet option. It communicates with the PC using an X540-T2 10GbE card. A full \qty{72}{\giga \byte} camera acquisition can be transferred to the computer in \qty{\sim 2}{\minute}.

\subsection{Refractive index reconstruction}
FS-ODT reconstructions were carried out using custom Python code~\cite{mcsim} running on Python 3.11.4 or 3.11.5 and CUDA toolkit 11.8.0. We implemented light scattering forward models and proximal gradient algorithms, harnessing CuPy for GPU acceleration and Dask for parallelization and orchestration. We rely on the RAPIDS cuCIM implementation of the total variation proximal operator, which is modeled on the scikit-image implementation, \texttt{denoise\_tv\_chambolle}. For weighted phase unwrapping during computing the Rytov field, we adapted the approach of~\cite{ghigliaRobustTwodimensionalWeighted1994} to run on the GPU. 

Reconstructions were performed on either the experimental control computer described above, or a custom-built computer running Ubuntu 20.04.6 LTS with an X99S SLI Plus motherboard (MSI), an i7-5820K CPU (Intel), \qty{125}{\giga \byte} RAM, and an RTX A6000 GPU (NVidia) with \qty{48}{\giga \byte} of memory.

\subsection{Preparation of \emph{Tetrahymena} samples}
\emph{Tetrahymena} were cultured at room temperature in media containing \qty{2}{\percent} proteose peptone, \qty{0.2}{\percent} yeast extract, \qty{0.012}{\milli \molar} \ch{FeCl3}, \qty{0.2}{\percent} glucose, \qty{100}{\unit / \milli \liter} penicillin, \qty{100}{\milli \gram / \milli \liter} streptomycin, and \qty{0.25}{\milli \gram / \milli \liter} Amphotericin B. \emph{Tetrahymena} were passaged to fresh media every 3-4 days. \emph{Tetrahymena} were stained as described previously at room temperature unless noted~\cite{galatiDisApdependentStriatedFiber2014}. Briefly, \num{\sim 2e6} mid-log cells were washed with \qty{10}{\milli \molar} Tris pH \num{7.5}, permeabilized in \qty{0.25}{\percent} TX-100 in PHEM for \qty{30}{\second} (\qty{60}{\milli \molar} PIPES, \qty{25}{\milli \molar} HEPES, \qty{10}{\milli \molar} EGTA, \qty{4}{\milli \molar} \ch{MgSO4}), fixed in \qty{1}{\percent} paraformaldehyde in PHEM for \qty{15}{\minute}, blocked in \qty{3}{\percent} BSA in PBS-T for \qty{30}{\minute} (\qty{0.01}{\percent} Tween-20, \qty{130}{\milli \molar} \ch{NaCl}, \qty{2}{\milli \molar} \ch{KCl}, \qty{8}{\milli \molar} \ch{Na2HPO4}, \qty{2}{\milli \molar} \ch{KH2PO4}, \qty{10}{\milli \molar} EGTA, \qty{2}{\milli \molar} \ch{MgCl2}, pH 7.5), stained with primary antibodies diluted in \qty{3}{\percent} BSA in PBS-T at \qty{4}{\celsius} overnight (glutamylated tubulin 1:1000, GT335, Adipogen, AG-20B-0020-C100 and centrin 1:1000, 20H5, MilliporeSigma, 04-1624), stained with secondary antibodies diluted in \qty{3}{\percent} BSA in PBS-T for 1 hour (1:1000 goat anti-mouse IgG2a Alexa Fluor 488, Invitrogen, A-21131 and 1:1000 goat anti-mouse IgG1 Rhoadmine Red X, Jackson Immuno, 115-005-205), and counterstained with \qty{1}{\milli \gram / \milli \liter} DAPI for \qty{5}{\minute} . Samples were washed with PBS three times after fixation, primary antibody incubation, and secondary antibody incubation. Each wash lasted \qty{5}{\minute} and all centrifugations were carried out for \qty{5}{\minute} at $250 \times g$ in a swinging bucket centrifuge.

For the samples in Fig.~\ref{fig:schematic}b, we placed \qty{\sim 100}{\micro \liter} of fixed \emph{Tetrahymena} cells in PBS in a \qty{40}{\milli \meter} diameter cell culture dish (FluoroDish FD5040). For the samples in Fig.~\ref{fig:schematic}c, \qty{10}{\micro \liter} of fixed \emph{Tetrahymena} cells in PBS were placed on a round \#\num{1.5} coverslip of diameter \qty{40}{\milli \meter}. A square \#\num{1} coverslip with \qty{25}{\milli \meter} sides was dropped on top and the chamber was sealed with epoxy (Devcon 2 Ton Epoxy, GLU-735.90). The electric field data was binned by a factor of \num{2} to reduce the memory required during reconstruction.

\subsection{\qty{10}{\micro \meter} polystyrene microspheres}
PMS of diameter \qty{10}{\micro \meter} (Thermofish Fluospheres F8836) were sonicated for \qty{20}{\minute} and then diluted by a factor of \num{10} in immersion oil (1.04699.0500 MilliporeSigma) with RI in the range \numrange{1.515}{1.517}. We spread \qty{20}{\micro \liter} of the resulting emulsion on a \#\num{1.5} coverslip of diameter \qty{40}{\milli \meter} with a pipette tip and left it uncovered overnight for the water to evaporate. Then we dropped a \#1 square \qtyproduct{25x25}{\milli\m} coverslip on top and sealed the chamber with epoxy (Devcon 2 Ton Epoxy, GLU-735.90). After the epoxy dried, the sample was placed on the microscope, and a few drops of water were placed on the top coverslip to facilitate imaging with a water dipping objective.

\subsection{Preparation of diffusing microspheres in a water-glycerol mixture}
PMS of diameter $2R = \qty{1}{\micro \meter}$ (ThermoFisher Fluospheres F8823) of weight/volume \qty{0.02}{\gram \per \milli \liter} were first sonicated for \qty{20}{\minute} and then diluted by a factor of \num{10} in MQ water. \qty{10}{\micro \liter} of this dilution was combined with \qty{40}{\micro \liter} of MQ water and \qty{50}{\micro \liter} of glycerol. \qty{10}{\micro \liter} of this mixture was placed on a round \#\num{1.5} coverslip of diameter \qty{40}{\milli \meter}. A square \#\num{1} coverslip with \qty{25}{\milli \meter} sides was dropped on top, and the chamber was sealed with epoxy (Devcon 2 Ton Epoxy, GLU-735.90).

This experiment was performed on an earlier version of the apparatus. The detection objective was a $50 \times$ long-working distance air objective with $\na = \num{0.55}$ (Mitutoyu 378-805-3) and \qty{200}{\milli \meter} focal length tube lens (Thorlabs ACT508-200-A-ML). The camera was a Prime BSI Express (Teledyne Photometrics) with \qty{6.5}{\micro \meter} pixels and \qty{1.8}{\electron} RMS read noise. No beam expander was used after the detection tube lens. The effective pixel size was \qty{0.130}{\micro \meter}. The frame rate was limited by the readout time of the camera.

The viscosity of this mixture is $\eta = \qty{0.0076}{\pascal \second}$, and the density is $\rho_\text{solvent} = \qty{1.13}{\gram / \milli \liter}$ at $T = \qty{22.5}{\celsius}$. The expected diffusion coefficient far from the wall is $D_o = k_B T / 6\pi \eta R = \qty{0.0579}{\micro \meter^2 / \second}$. The measured average diffusion coefficient is \qty{0.0536}{\micro \meter^2 / \second}. Deviations may come from differences in temperature and magnification from nominal values and the hydrodynamic interactions with the wall. The density of polystyrene is $\rho_\text{bead} = \qty{1.05}{\gram / \milli \liter}$ and therefore the microspheres are buoyant in this mixture. We observe that the height distribution of the microspheres matches a Boltzmann distribution with the expected effective mass $m = \frac{4}{3} \pi R^3 (\rho_\text{bead} - \rho_\text{solvent})$.

After reconstructing the sample RI, we identified and localized the microspheres using a custom Python package, \texttt{localize-psf} \url{https://github.com/QI2lab/localize-psf}. To identify candidate microspheres, we first applied a difference-of-Gaussian filter to suppress noise and background, then applied a maximum filter and selected pixels where the initial and maximum filtered images had the same values. We keep only points above a threshold. Next, we fit an ROI of size \qtyproduct{4 x 2 x 2}{\micro \meter} around each candidate to a 3D Gaussian and only kept spots where the fit parameters fell within certain bounds. After localizing the microspheres at all time points, we tracked them using \texttt{trackpy} v0.6.1, \url{http://soft-matter.github.io/trackpy}. To mitigate the effect of localization errors, we computed the mean-square displacements using every \num{8}th frame.

\subsection{Preparation of \emph{E. coli} and microspheres}
A culture of \emph{Escherichia coli} strain RP437, which is wild type for motility, was inoculated from a single colony and incubated overnight at \qty{37}{\celsius} in Lysogeny Broth (\qty{10}{\gram \per \liter} tryptone, \qty{5}{\gram \per \liter} yeast extract, \qty{10}{\gram \per \liter} \ch{NaCl}). Cultures were diluted 1:100 in T-broth (\qty{10}{\gram \per \liter} tryptone, \qty{10}{\gram \per \liter} \ch{NaCl}), and incubated at \qty{33}{\celsius} while shaking at \qty{200}{\rpm}, until reaching an $O.D_{600}$ of \num{0.5}. Cells were then centrifuged at $1,200 \times g$  for \qty{7}{\minute}, washed, and resuspended in motility buffer (\qty{10}{\milli \molar} \ch{K2HPO4}, \qty{0.1}{\milli \molar} EDTA, pH 7.5). The mobility buffer's RI was \num{1.333} as measured using a refractometer (Kr{\"u}ss HR 901). The bacterial concentration was estimated to be \qty{1.2e8}{1/\milli \liter} using a hemacytometer (Fisher Scientific 0267151B).

PMS of diameter \qty{0.5}{\micro \meter} (ThermoFisher Fluospheres F8813) of weight/volume \qty{0.02}{\gram \per \milli \liter} were first sonicated for \qty{20}{\minute} and then diluted by a factor of \num{3e2} with the motility buffer and \emph{E. coli} mixture. Coverslips were cleaned with EtOH. A \qty{120}{\micro \meter} tall chamber was prepared by placing a secure seal spacer (Electron Microscopy Science Cat \#70327-20S) on a round \#\num{1.5} coverslip of diameter \qty{40}{\milli \meter}. \qty{\sim 40}{\micro \liter} of solution were placed in this chamber, and a square \#\num{1} coverslip with \qty{25}{\milli \meter} sides was placed on top to close the chamber.

To quantify the bacteria swimming and identify the wobble angle, we trained a classifier to distinguish between \emph{E. coli}, tracer particles, and background pixels using LabKit~\cite{arzt_labkit_2022}. Next, we labelled individual bacteria and identified their main axis using principal component analysis using \texttt{scikit-image}, and tracked the bacteria with \texttt{trackpy}. To ensure we tracked only swimming bacteria, we exclude any tracks shorter than \num{50} frames or which move a total distance less than \qty{4}{\micro \meter}. To generate the smoothed velocity, we first smoothed the centroid position by computing a rolling average over a window of size \num{15} and then computed the velocity using a second-order difference method~\cite{liu_bacterial_2024}.

\subsection{Preparation of diffusing microspheres in water}
PMS of diameter \qty{1}{\micro \meter} (ThermoFisher Fluospheres F8823) were sonicated for \qty{20}{\min} and then \qty{3}{\micro \liter} were diluted with \qty{97}{\micro \liter} of MQ water. For the shorter chamber, \qty{10}{\micro \liter} of the dilution was pipetted on a \#1.5 coverslip of diameter \qty{40}{\milli \meter} and a square \#1 coverslip with \qty{25}{\milli \meter} sides was dropped on top. The chamber was sealed with epoxy (Devcon 2 Ton Epoxy, GLU-735.90). The height of the fluid chamber was estimated to be \qty{\sim 10}{\micro \meter} from a fluorescence z-stack. For the taller chamber, we placed a \qty{120}{\micro \meter} thick spacer (Electron Microscopy Sciences 70327-20S) on the round coverslip, pipetted \qty{40}{\micro \liter} of solution, and closed the chamber with the square coverslip. To achieve a DMD-limited frame rate, we cropped the camera chip to \numproduct{960 x 1040} pixels. We identified and tracked the PMS using the same procedure as for the PMS in the water-glycerol mixture.

\section*{Supporting Information}

\renewcommand{\thesection}{S\arabic{section}}
\renewcommand{\thesubsection}{\Alph{subsection}}
\renewcommand{\thefigure}{S\arabic{figure}} 
\renewcommand{\theequation}{S\arabic{equation}}

\setcounter{section}{0}
\setcounter{figure}{0}
\setcounter{equation}{0}

\section{FS-ODT multiplexing validation with Mie theory}
\begin{figure}[htb!]
	\centering
	\includegraphics[width=.8\linewidth]{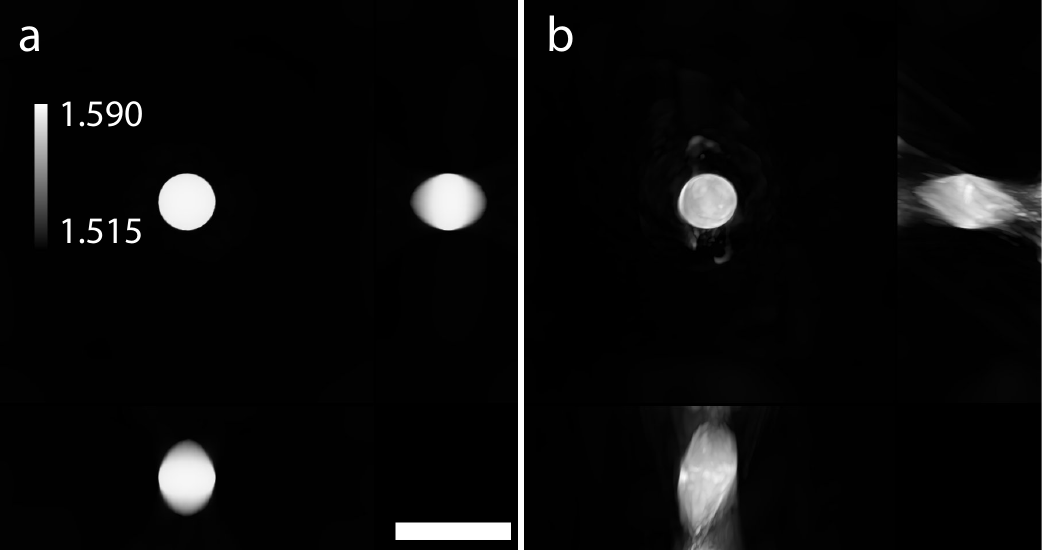}
	\caption{
    \textbf{Validating multiplexing with Mie theory}.
    \textbf{A}. Maximum intensity projections of refractive index reconstruction of a simulated \qty{10}{\micro \meter} diameter sphere using $10 \times$ multiplexing. The axial refractive index reconstruction is slightly distorted due to missing cone artifacts and the limited excitation and detection numerical aperture of \num{1}. Scale bar \qty{20}{\micro \meter}.
    \textbf{B}. Maximum intensity projections of refractive index reconstruction of a simulated \qty{10}{\micro \meter} diameter sphere without multiplexing. The limited angular information produces an inferior reconstruction.
    \label{fig:mie_reconstruction}
    }
\end{figure}

We validate our reconstruction approach by comparing with exact results using Mie theory~\cite{bohren_chapter4_1983}. The Mie theory solution to scattering from a spherical dielectric particle expresses the field as a sum of vector spherical harmonics. Each spherical harmonic is a product of sinusoids, spherical Bessel functions, and associated Legendre polynomials. Although many Python packages calculate the expansion coefficients, few compute the electric fields. We rely on miepython~\cite{miepython} to compute the expansion coefficients and developed GPU accelerated Python code to calculate the Mie fields. Our code relies on CuPy functions where possible. Since the second order spherical Bessel functions $y_n$ are not implemented in CuPy, we wrote a CUDA kernel which computes $y_n$ using downward recursion and Miller's algorithm~\cite{olver_error_1964}. To avoid the need to rerun this computation for $n = 1, ... N$, our routine provides the full sequence $y_1, ..., y_n$. For an image size of \numproduct{700 x 650} pixels, the CPU implementation runs in \qty{\sim 120}{\second} while the GPU runs in \qty{\sim 0.4}{\second}.

Using Mie theory simulations, we validate that our multipled ODT patterns provide considerably more information than non-multiplexed patterns and consequently produce superior reconstructions, compared with non-multiplexed patterns when the number of patterns is kept fixed. We compare results for \num{15} images using $10 \times$ multiplexing and without multiplexing. To generate simulated images, we first compute the Mie electric fields using the expansion discussed above for a sphere of diameter \qty{10}{\micro \meter} with refractive index of \num{1.59} in media with background refractive index \num{1.515}, chosen to match a PMS microspheres in immersion oil. Our routine returns both the plane waves fields and the scattered fields. We apply a phase shift to each field, and then sum them to simulate multiplexing. We interefere this electric field with a reference plane wave. Then we produced detected fields by applying photon shot noise, camera gain and offset, and camera readout noise. We keep the maximum photon number fixed at \num{\sim 40000} for the final images. We reconstruct these simulated images following the same procedure as for the experimental data. 

We find that the multiplexed images (fig.~\ref{fig:mie_reconstruction}A) produce superior reconstructions to the non-multiplexed images (fig.~\ref{fig:mie_reconstruction}B). Note that unlike the comparison in fig.~\ref{fig:10um_bead}, the number of images has been kept fixed rather than the number of plane waves. Due to the limited number of plane waves used with the non-multiplexed images, significant reconstruction artifacts are present in the inferred refractive index. Therefore, for a fixed time budget, we expect that multiplexed images provide a superior reconstruction. 

\section{Optical diffraction tomography\label{section:odt_details}}
Optical diffraction tomography was performed using a Mach-Zender interferometer, part of a bespoke multimodal microscope with quantitative phase imaging (FS-ODT) and multicolor fluorescence superresolution microscopy (structured illumination) capabilities. Portions of this microscope are described in previous work~\cite{Brown2021}. Up to \qty{80}{\milli \watt} of \qty{785}{\nano \meter} light with coherence length of \qty{\sim 50}{\meter} is generated using a volume-holographic-grating (VHG) stabilized laser (Thorlabs FPV785P). This light is divided using a polarizing beam splitter, and the two paths are coupled into separate \qty{1}{\meter} long polarization-maintaining fibers (Thorlabs PM780-HP). One fiber is used to generate the ODT excitation light, and the other is used to generate the reference beam for the off-axis holography. 

The reference beam is collimated with a molded aspheric lens of focal length \qty{13.86}{\milli \meter} (Thorlabs C560TME-B) and beam expanded with lenses of focal length \qty{40}{\milli \meter} (Thorlabs AC254-040-B-ML) and \qty{300}{\milli \meter} (Thorlabs AC508-300-AB-ML). It is combined with the reference beam on a d-mirror (Thorlabs BBD1-E03) near the focal plane of the \qty{40}{\milli \meter} lens. 

The excitation light is collimated with a molded aspheric lens of focal length \qty{18.4}{\milli \meter} (Thorlabs A280TM-B) and beam expanded by lenses with focal length \qty{30}{\milli \meter} (Thorlabs AC254-030-AB-ML) and \qty{125}{\milli \meter} (Thorlabs LA1986-B-ML), resulting in a Gaussian beam with waist \qty{\sim 7}{\milli \meter} which is incident on the DMD.

The ODT patterns are generated from diffraction off a DMD (Texas Instruments DLP6500) with \qty{7.56}{\micro \meter} pitch and \numproduct{1920 x 1080} mirrors. We have discussed the details of our DMD geometry elsewhere~\cite{Brown2021}. Let $\uvec{x}$ point along the long axes of the mirror grid and $\uvec{y}$ point along the short axis, both in the plane of the DMD face. Let $\uvec{p} = (\uvec{x} + \uvec{y}) / \sqrt{2}$, $\uvec{m} = (\uvec{x} - \uvec{y}) / \sqrt{2}$, and $\uvec{z}$ be the normal vector of the DMD surface pointing outwards. The DMD mirrors swivel about $\uvec{p}$ the axis and can be in two binary states, either $\gamma_\pm = \qty{\pm 12}{\degree}$, which we will call the $+$ and $-$ states. The DMD chip is rotated so that the optical table normal points along $\uvec{p}$, ensuring that the principle diffraction occurs in the $mz$ plane.

The initial DMD geometry was designed to enable 3 color SIM with excitation wavelength \qty{465}{\nano \meter}, \qty{532}{\nano \meter}, and \qty{635}{\nano \meter}. For these three colors to all roughly meet the blaze condition, the DMD face normal $\uvec{z}$ makes an angle of $\theta_d \sim \qty{-21.2}{\degree}$
with the optical axis. In the DMD coordinate system, the optical axis points along the $\sin \theta_d \uvec{m} + \cos \theta_d \uvec{z}$.

To achieve high-efficiency ODT, we align the \qty{785}{\nano \meter} excitation light to approximately satisfy the blaze condition for the $(n_x, n_y) = (-3, 3)$ diffraction order when the mirrors are in the $-$ state. The excitation light is incident on at an angle of approximately $\theta_e = \qty{6.64}{\degree}$ so that $\uvec{v}_e = \sin \theta_e \vec{m} - \cos \theta_e \uvec{z}$. The light diffracted into $(-3, 3)$ travels along the $\theta_o = \qty{-18.96}{\degree}$ along $\uvec{v}_o = \sin \theta_o \vec{m} + \cos \theta_o \uvec{z}$.

As discussed in the main text, we apply an additional ``carrier frequency'' to our patterns of frequency $\vec{f}_c = \frac{1}{4} \uvec{x} - \frac{1}{4} \uvec{y}  \ 1/\text{mirror}$. This produces additional diffraction about the $(-3, 3)$, and we have designed our system such that the $(-3, 3) + (1/4, -1/4)$ diffraction order travels along the optical axis and is nearly blazed. The perfectly blazed output direction for the $-$ mirrors is $\sin \theta_b \uvec{p} + \cos \theta_b \uvec{z}$ for $\theta_b = \qty{-17.36}{\degree}$. An aperture blocks all other diffraction orders. This specific carrier frequency is chosen to displace the beam as far as possible from orders of the form $(n/2, m/2)$, as we see significant diffraction due to the $+$ mirror states. Furthermore, this avoids diffraction along the $\uvec{x}$ and $\uvec{y}$ axes coming from a row of DMD mirrors beyond the active chip, which are fixed in the $-$ state.

The DMD is in a conjugate plane to the back focal plane (Fourier plane) of the excitation and detection objectives. After light diffracts off of the DMD, it is relayed by a pair of imaging systems, the first using lenses of focal length \qty{200}{\milli \meter} (Nikon MXA20696) and \qty{100}{\milli \meter} (Thorlabs AC508-100-A-ML), and the second using lenses of \qty{400}{\milli \meter} (Thorlabs AC508-400-A-ML) and \qty{300}{\milli \meter} (Thorlabs AC508-300-A-ML). After the Nikon tube lens, the NIR light is separated from the visible light with a dichroic mirror (Semrock FF750-SDi02-25x36). After the \qty{100}{\milli \meter} achromat, they are recombined using a second identical dichroic mirror. The two dichroic mirrors are arranged at right angles to each other to avoid the differential $s$-- and $p$--phase shifts from affecting the polarization of the visible light~\cite{LuWalther2015}. We align the polarization orthogonal to the table surface to avoid similar polarization degradation of the ODT beam by the epifluorescence dichroic for the fluorescence modality.

The ODT light is focused with an oil immersion objective (Olympus UPlanFL N 100x NA 1.3), interacts with the sample, and is collected using a water objective (Olympus LUMPLFLN60XW) and a \qty{180}{\milli \meter} tube lens (Thorlabs AC508-180-AB-ML). The image is then magnified by a factor of \num{3} using a relay composed of a \qty{100}{\milli \meter} (Thorlabs AC508-100-B-ML) and \qty{300}{\milli \meter} (Thorlabs AC508-300-AB-ML) lenses. The light is imaged onto a Phantom camera (VEO-1010L-72G-M) which has \numproduct{1280 x 960} pixels, pixel size \qty{18}{\micro \meter}, quantum efficiency \qty{\sim 51}{\percent} at \qty{785}{\nano \meter}, read-noise \qty{10.5}{\electron} RMS and gain of \qty{0.39}{\adu / \electron}, measured using the approach of~\cite{Huang2013}. The effective pixel size is \qty{0.1}{\micro \meter}. The maximum frame rate for the full chip is \qty{8420}{\frame / \second}, but it can be increased by cropping the chip.

\begin{figure}[htb!]
	\centering
	\includegraphics[width=\linewidth]{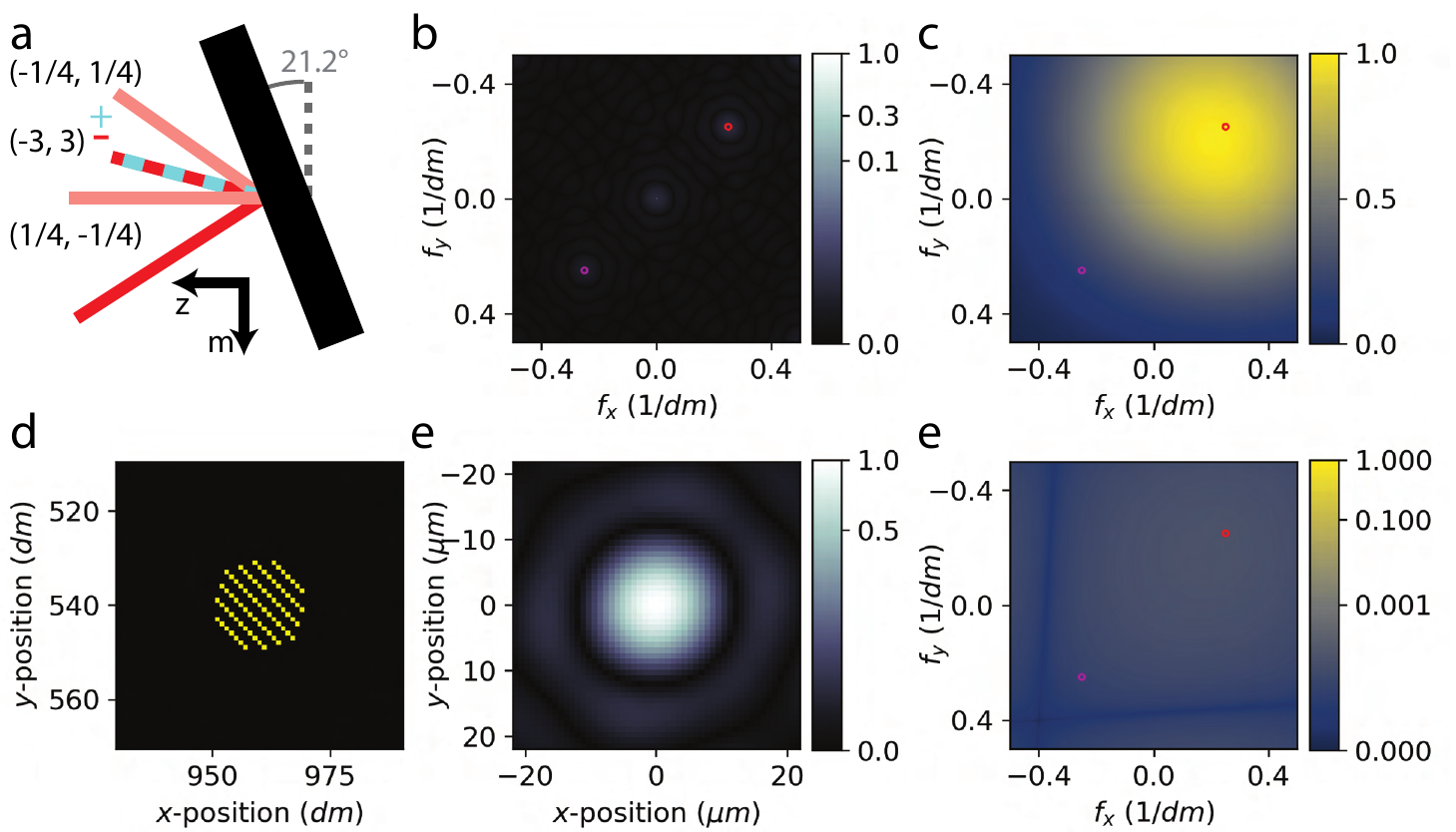}
	\caption{
    \textbf{DMD geometry and pattern simulation}.
    \textbf{A}. DMD geometry showing the incident NIR light (red) and the main diffraction orders for the $+$ (cyan) and $-$ (red) mirrors and the positive and negative diffraction at the carrier frequencies (pale red)
    \textbf{B}. Magnitude squared of the diffracted electric field versus frequency, $|E(f)|^2$. The diffracted weight is concentrated around the positive (red) and negative (magenta) carrier frequency diffraction orders 
    \textbf{C}. Blaze envelope for $-$ mirrors. 
    \textbf{D}. Sample DMD pattern with $R = \qty{10}{\mirror}$ show mirrors in the $-$ state (red) and the $+$ state (cyan)
    \textbf{E}. Pattern in sample plane with measured waist $w_o = \qty{9.3}{\micro \meter}$
    \textbf{F}. Blaze envelope for the $+$ mirrors. The strength of this diffraction is suppressed by \num{\sim e-5} compared with the $-$ mirrors.
    \label{fig:dmd_geometry}
    }
\end{figure}

\subsection{DMD efficiency\label{section:dmd_eff}}
The expected peak DMD diffraction efficiency into any order $(n_x, n_y)$ is \qty{\sim 50}{\percent}, limited by the reflective of the DMD mirrors, the fraction of the chip the mirrors cover, the transmissivity of the DMD window, and diffraction physics. Using the carrier frequency, we expect that efficiency into the $\pm 1$ carrier orders are each \qty{\sim 25}{\percent} of the power in the $0$th order. We expect the power in $0$th order is \qty{\sim 18}{\percent} of the power that would be diffracted if the mirrors were all in the $-$ state. Taken together, we estimate the DMD diffraction efficiency into the desired order is \qty{\sim 2}{\percent}.

The fiber coupling efficiency is \qty{\sim 50}{\percent}. Additionally, since the fluorescence modality of our microscope operates at visible wavelengths, most of the optical coatings are optimized for visible light. Thus, only \qty{\sim 75}{\percent} of light diffracted from the DMD reaches the objectives. The two objective lenses have a combined transmissivity of \qty{\sim 50}{\percent}. The efficiency from the DMD to the camera is thus \qty{\sim 30}{\percent},

As only a small fraction of the DMD mirrors are used to generate a plane wave, this further reduces the efficiency. We adjust the magnification between the back focal plane and the DMD so that the pupil radius is approximately the same size as the DMD along its narrow dimension. As the magnification factor is $M = \num{0.625}$, the detection objective pupil radius is $R_p = \na f / 0.625 = \qty{4.8}{\milli \meter}$ at the DMD. For a spot pattern of radius $R$ and assuming the laser power is distributed uniformly over the pupil, we expect the number of photons per second that strike the camera is
\begin{equation}
\frac{\qty{80}{\milli \watt}}{h c / \lambda} \times \frac{\pi R^2}{\pi R^2_p} \times \left( 0.5 \times 0.02 \times 0.3 \right) \approx \qty{2e11}{\photon / \second}.
\end{equation}
for $R = 10 \times \qty{7.56}{\micro \meter}$. Here, we divide the terms into incident power, geometric efficiency, and transmission/diffraction efficiency.

For patterns of radius \num{10} mirrors, our simulations show that the beam waist in the imaging plane is \qty{\sim 10}{\micro \meter} (Fig.~\ref{fig:dmd_geometry}E), and putting this all together, for an imaging time of \qty{100}{\micro \second} we expect to collect about
\begin{equation}
    N = \qty{2e11}{\photon / \second} \times \frac{\left( \qty{20}{\micro \meter} \times \frac{1}{180} \right)^2}{\pi \left( \qty{10}{\micro \meter} \right)^2} \times \qty{100}{\micro \second} \times \qe \sim \qty{400}{\photon}
\end{equation}
per pixel, where the quantum efficiency of the camera is $\qe \sim \qty{50}{\percent}$ at \qty{785}{\nano \meter}.

\subsection{Pattern fidelity\label{section:fidelity}}
Previous ODT approaches using binary DMD patterns have resulted in lower-quality reconstruction than gray-scale approaches due to the unwanted additional diffraction orders introduced by the DMD's square binary pixels. Our previous structured illumination microscopy work addressed similar issues~\cite{Brown2021}. In both cases, these spurious diffraction orders arise when using large-scale periodic patterns covering the face of the DMD. In this case, our small spot patterns lead to much smaller contributions from these orders, which are mitigated by Fourier broadening (Fig.~\ref{fig:dmd_geometry}B).

\subsection{DMD non-planarity\label{section:non_planarity}}
Unlike in many DMD imaging applications, we do not place the DMD orthogonal to the optical axis (Fig.~\ref{fig:dmd_geometry}A). This compromise improves the diffraction efficiency by satisfying the blaze condition. In our geometry, the effect of this tilt is minor. 

The tilt introduces a shift in the focus of the plane waves across the DMD face. At the pupil radius, the DMD $z$-shift is at most $R_p \sin \theta_D \sim \qty{1.73}{\milli \meter}$. As the $z$-magnification is $M_z = M^2 = \num{0.39}$ the shift in the objective back-focal plane is \qty{0.67}{\milli \meter}. The shift must be compared with the depth-of-focus of the beam in the back focal plane, which we estimate using the Rayleigh range. For a plane wave with $w_o = \qty{10}{\micro \meter}$ in the focal plane, the waist in the objective BFP is \qty{\sim 37}{\micro \meter} corresponding to a Rayleigh range of $\pi w_o^2 / \lambda \sim \qty{5.5}{\milli \meter}$, which is one order of magnitude larger than the focal shift.

The tilt also introduces deformations in the transformation between the position on the DMD face and the frequency of the beam. For example, there is some shear, and a ring pattern on the DMD maps to an oval in frequency space. The precise effects can be calculated using the approach of~\cite{Brown2021} described in section 3 of the supplemental methods.

\subsection{System stability\label{section:stability}}
\begin{figure}[htb!]
	\centering
	\includegraphics[width=\linewidth]{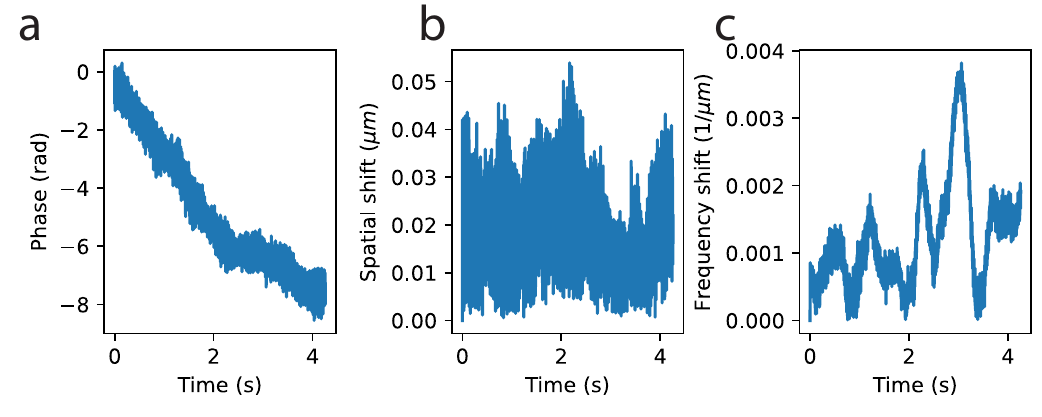}
	\caption{
    \textbf{System stability}.
    \textbf{A}. Exemplary phase stability plot from the first pattern from the \qty{10}{\micro \meter} bead data. Phases are unwrapped
    \textbf{B}. Exemplary pattern position stability for the same pattern as A.
    \textbf{C}. Exemplary pattern frequency stability for one frequency in the pattern in A. 
    \label{fig:stability}
    }
\end{figure}

Due to the relatively long beam path ($> \qty{2}{\meter}$) used in our multimodal microscope setup, we observe several sources of instability in our system and ODT patterns, which we correct for computationally during ODT reconstruction. Specifically, we correct for (1) phase drift between the reference arm and the imaging arm, (2) frequency instability of the ODT patterns, and (3) position instability of the ODT patterns.

To correct for (1), we determine the complex factor relating the image electric fields to a single background electric field using a least-squares fit. To correct for (2), we determine the location of the Fourier peaks versus time by fitting the Fourier transform of the hologram image to a Gaussian in the vicinity of each peak. To correct for (3), we register images using the Fourier transform of the absolute value of the electric field. Empirically, we find that using the absolute value of the electric field is superior to using either the Fourier transform of the intensity or the Fourier transform of the electric field. The improvement is likely due to the rejection of background fluctuations that do not involve the interference pattern and taking the absolute value removes the need to consider phase fluctuations of the electric field.

The stability of our system over \qty{\sim 4}{\second} is shown in Fig.~\ref{fig:stability}. Over this short time span the relative phase between the imaging and reference beam drifts by $\sim 2 \pi$ radians, the spatial position drifts by \qty{< 50}{\nano \meter}, and the beam frequency drifts by \qty{< 4e-3}{\micro \meter^{-1}}. Note that the Fourier space pixel sizes are $df_x = \qty{7.81e-3}{\micro \meter^{-1}}$ and $df_y = \qty{10.42e-3}{\micro \meter^{-1}}$, so the frequency drift is sub-pixel.

\section{Iterative reconstruction with proximal gradient methods\label{section:reconstruction}}
As FS-ODT multiplexing introduces a more challenging computational image reconstruction problem requiring new approaches we briefly discuss some commonly used reconstruction approaches. In 1969, Wolf introduced optical diffraction tomography using a Born approximation formulation \cite{Wolf1969}. In this formulation, the forward model describing the electric field after interacting with a given RI distribution is linear. This linearity makes solving the inverse problem, i.e. inferring the RI based on the observed electric fields, particularly simple. Devaney realized that the Rytov approximation is better suited for biological sample \cite{Devaney1981}, and various technical improvements have expanded the range of validity and quality of reconstructions \cite{Kostencka2015, Krauze2018, Devaney1982b, Kim2013, kim2013a, Mueller2015b, Kostencka2016, Kostencka2016a}. However, imaging thicker and higher-contrast samples inevitably introduces multiple scattering, which the linear Born and Rytov approximations do not include. To address multiple scattering, a variety of multi-slice reconstruction approaches have been developed that rely on the beam-propagation model (BPM) \cite{Tian2015b, Kamilov2015, Kamilov2016, Lim2018, Chowdhury2019}. However, the BPM entails the paraxial approximation, motivating the development of more accurate forward models, including the split-step non-paraxial (SSNP) model~\cite{Lim2019}, more sophisticated Born approximation approaches \cite{Kamilov2016a, Chen2020, Lee2022}, and HyPM \cite{moserEfficientAccurateIntensity2023}. Most approaches consider forward scattering only, which is implicit in the layer-by-layer multi-slice forward models. Other approaches account for backscattering using the Lippman-Schwinger equation \cite{Soubies2017, Pham2018,Liu2018a}, but these are generally more computationally expensive than multi-slice models. Machine learning approaches are increasingly employed to accelerate and denoise RI reconstruction~\cite{matlockMultiplescatteringSimulatortrainedNeural2023}. In this work, we primarily rely on the BPM or the SSNP combined with an initial guess generated using a demultiplexed low-resolution Rytov approximation approach to infer RI information from FS-ODT data.

We treat refractive index reconstruction as a regularized minimization problem and solve it using the fast iterative shrinkage-thresholding algorithm (FISTA)~\cite{Beck2009}. As usual in FISTA, we attempt to minimize a function which is the sum of two terms, a loss function and a regularization function $\mathcal{L}(x) + g(x)$, where $\mathcal{L}$ is differentiable and has a Lipschitz-continuous gradient with Lipschitz constant $L$, and $g$ is convex. We iteratively update the proposed solution, $x_t$, starting from $t=0$, $q_0 = 1$, the step-size $\gamma$, and a starting guess $x_o$
\begin{align*}
    y_t &= \prox_{\gamma_t} \left[x_{t-1} - \gamma \nabla \mathcal{L}(x_{t-1}) \right]\\
    q_t &= \frac{1 + \sqrt{1 + 4 q_{t-1}^2}}{2}\\
    x_{t + 1} &= y_t + \frac{q_{t-1} - 1}{q_t} \left(y_t - y_{t-1} \right).
\end{align*}
In the last step the convergence is accelerated by including a momentum term and the structure of $q_t$ is chosen to change the convergence from $\mathcal{O}(1/t)$ to $\mathcal{O}(1/t^2)$.

The proximal operator for $g$,
\begin{equation}
        \prox_\gamma(z) = \argmin_x \left \{g(x) + \frac{1}{2 \gamma} \left \vert x - z \right \vert^2 \right\},
\end{equation}
determines a new object which is near to the initial value but better satisfies the regularization. When an explicit form of or fast algorithm for computing the proximal operator is known, this process is efficient, as in the case of total variation~\cite{Chambolle2004}, $\ell_1$, or $\ell_2$ norms. In this work we choose a regularization function which enforces smoothness and sparsity
\begin{equation}
    g(x) = \tau_\text{tv} TV(x) + \tau_{\ell_1} \lVert x \rVert_1.
\end{equation}

The total loss, $\mathcal{L}(x_t) + g(x_t)$, is non-increasing when the step-size $\gamma \in [0, 2/L]$. If the Lipschitz constant is not known, a line-search strategy can determine the step-size at each iteration by reducing an initial step-size until the Lipschitz condition is locally satisfied (algorithm~\ref{alg:line_search}). This requires additional calculation of $\mathcal{L}(y)$, $\mathcal{L}(x)$, and $\nabla \mathcal{L}(y)$.
\begin{algorithm}
\KwData{Proposed object $x$}
\KwData{Initial step-size $\gamma_o$}
\KwData{Step-size adjuster $\alpha < 1$}
$\gamma \gets \gamma_o$\;
$y\gets \prox \left[x - \gamma \nabla \mathcal{L}(x) \right]$\;
\While{$\mathcal{L}(y) > \mathcal{L}(x) + \nabla \mathcal{L}(x) \cdot (y - x) + \frac{1}{2 \gamma} \lVert y - x \rVert^2_2$}
{
    $\gamma \gets \alpha \gamma$\;
    $y\gets \prox \left[x - \gamma \nabla \mathcal{L}(x) \right]$\;
}
\caption{Line-search algorithm for setting the step-size}\label{alg:line_search}
\end{algorithm}

In ODT, we measure a sequence of electric fields $\vec{\Psi}^{(i, d)}$, where $i = 1, ..., M$ indexes the incident angles, derived from off-axis holography and define our loss function by
\begin{align}
    \mathcal{L} &= f \mathcal{L}_E + (1 - f) \mathcal{L}_I \label{eq:loss_fn_mix}\\
    \mathcal{L}_E(n) &= \frac{1}{2 M} \sum_{i=1}^M \left \Vert \vec{\Psi}^{(i)} - \vec{\Psi}^{(i, d)} \right \Vert^2_2 \label{eq:loss_fn}\\
    \mathcal{L}_I(n) &= \frac{1}{2 M} \sum_{i=1}^M \sum_{j=1}^N \left \vert \left | \Psi_j^{(i)} \right | - \left | \Psi_j^{(i, d)} \right| \right \vert^2 \label{eq:loss_fn_int}
\end{align}
where $f \in [0, 1]$ describes the relative strength of the phase-sensitive ($\mathcal{L}_E$) and phase-insensitive ($\mathcal{L}_I$) components of the loss function, $\vec{\Psi}^{(i)}(n)$ is the forward model describing the predicted electric field as a function of $n$, $\Vert \cdot \Vert_2$ is the $\ell_2$ norm, and $N$ is the number of pixels in a single electric field. Note that many different loss functions can be defined for the phase-insensitive portion, and in fact the most natural one might seem to involve $|\Psi|^2$, however previous work has demonstrated better convergence using eq.~\ref{eq:loss_fn_int}~\cite{Yeh2015}.

To perform FISTA, we require the gradient of the loss function with respect to the refractive index. The loss function, $\mathcal{L}: \mathbb{R}^N \times \mathbb{R}^N \to \mathbb{R}$, can be interpreted either using the real and imaginary parts of the refractive index as components of a vector, or as complex numbers. Since $\mathcal{L}$ is not holomorphic, if we regard the domain as $\mathbb{C}^N$ we must compute the gradient using Wirtinger derivatives~\cite{zhuHighfidelityIntensityDiffraction2022}, defined by $\partial_z = \frac{1}{2} \left( \partial_{z'} - i \partial_{z''} \right)$ and $\partial_{z^*} = \frac{1}{2} \left( \partial_{z'} + i \partial_{z''} \right)$, where $z'$ and $z''$ are the real and imaginary parts of $z$ respectively. In this formalism, the gradient used in FISTA is $\nabla = 2 \partial_{z^*}$. Computing this derivative, which only acts directly on the forward model, we find (suppressing the beam angle index)
\begin{align}
   \nabla_{n_b(z_l)} \mathcal{L} &= \sum_a \frac{\partial \mathcal{L}}{\partial \Psi_a^*} \frac{\partial \Psi_a^*}{\partial n_b(z_l)} \label{eq:loss_fn_grad}\\
     \partial_{\vec{\Psi}^*} \mathcal{L}_E &= \frac{1}{N} \left[\vec{\Psi} - \vec{\Psi}^{(d)} \right]\\
 \partial_{\vec{\Psi}^*} \mathcal{L}_I &= \frac{1}{N} \left[ \left | \vec{\Psi} \right | - \left | \vec{\Psi}^{(d)} \right| \right] \odot \frac{\vec{\Psi}}{\left | \vec{\Psi} \right|}
\end{align}
where $b$ and $l$ index the $xy$- and $z$-position of the voxels respectively and we have used the chain rule $\partial_{n^*} \left( \mathcal{L} \circ \Psi \right) = \partial_{n^*} \mathcal{L} \left(\partial_n \Psi \right)^* + \partial_n \mathcal{L} \left( \partial_{n^*} \Psi \right)$ and the fact $\partial_{n^*} \Psi = 0$.

Due to the FS-ODT pattern generation strategy, our illumination beams typically do not cover the entire field of view and exhibit decreasing intensity near the edges. In regions with less intensity, the influence of the loss function decreases and the effect of the regularization increases. Therefore, to achieve a higher quality reconstruction over a larger field of view it is helpful to normalize the loss function by the electric field. We optionally replace the loss function with
\begin{align}
    \mathcal{L}_E &= \frac{1}{2 M} \sum_{i=1}^M \sum_{j=1}^N \frac{\left \vert \Psi_j^{(i)} - \Psi_j^{(i, d)} \right \vert^2}{\left \vert \Psi_j^{(i, d)} \right \vert^2 + \alpha^2}\\
    \mathcal{L}_I &= \frac{1}{2 M} \sum_{i=1}^M \sum_{j=1}^N \frac{\left \vert \left | \Psi_j^{(i)} \right | - \left | \Psi_j^{(i, d)} \right| \right \vert^2}{\left | \Psi_j^{(i, d)} \right|^2 + \alpha^2},
\end{align}
where $\alpha$ is a regularization parameter that prevents division by small numbers where the electric field is noise dominated. The denominator does not depend on the forward model, so the loss function gradients need only be divided elementwise by the denominator to account for this change.

\subsection{Linear scattering models\label{section:linear_scattering}}
As usual we suppose light interacts with a spatially varying refractive index according to the scalar Helmholtz equation and work with phasors carrying $\exp (i \omega t)$ time dependence,
\begin{align}
    \left[\nabla^2 + k^2 n^2(\vec{r}) \right] E(\vec{r}) &= 0. \label{eq:helmholtz}
\end{align}
where $k = 2 \pi / \lambda n_o$.

Additionally we define the scattering potential $V$
\begin{align}
    V(\vec{r}) &= -\left(\frac{2\pi}{\lambda} \right)^2 \left[n^2(\vec{r}) - n_o^2 \right] \label{eq:scattering_pot}
\end{align}

We suppose that our sample is illuminated by a sequence of plane waves and the $i$th plane wave has frequency $\vec{f}^{(i)}$
\begin{align}
    E^{(i, o)} &= \exp \left[-i 2\pi \vec{f}^{(i)} \cdot \vec{r} \right]
\end{align}
where $2\pi \abs{\vec{f}^{(i)}} = k$.

In the Born approximation, valid when the cumulative phase shift of the beam is $\lesssim \pi / 4$ \cite{Slaney1984}, the 2D Fourier transform of the scattered electric field gives the scattering potential along a spherical cap in 3D Fourier space
\begin{align}
    \tilde{E}^{(i, s)}(f_x, f_y) &= \frac{1}{2i \times 2\pi f_z(f_x, f_y)} \tilde{V}(f_x - f^{(i)}_x, f_y - f^{(i)}_y, f_z - f^{(i)}_z) \; \text{(Born)}
\end{align}

The Rytov approximation is an alternate approach which is usually more accurate for biological samples. In this approximation, the scattering potential is related to the Rytov phase $\psi(\vec{r})$,
\begin{align}
    E^{(i, s)}(\vec{r}) &= E_o(\vec{r}) \left( e^{ \psi^{(i)}(\vec{r})} - 1 \right)\\
    \psi^{(i)}(\vec{r}) &= \log \left \vert \frac{E^{(i)}(\vec{r})}{E^{(i, bg)}(\vec{r})} \right \vert + i \ \text{unwrap} \left \{ \text{angle}\left[ E^{(i)}(\vec{r}) \right] - \text{angle} \left[ E^{(i, bg)}(\vec{r}) \right] \right \} \label{eq:rytov_phase}\\
    \tilde{\psi}^{(i)}(f_x - f_x^{(i)}, f_y - f_y^{(i)}) &= \frac{1}{2i \times 2 \pi f_z(f_x, f_y)} \tilde{V}(f_x - f_x^{(i)}, f_y - f_y^{(i)}, f_z - f_z^{(i)}) \; \text{(Rytov)}
\end{align}
The Rytov approximation is valid when $n^2(\vec{r}) - n_o^2 \gg \abs{\nabla \psi(\vec{r})}^2 k^2$~\cite{Devaney1981}.

In this case, it is more convenient to work with the scattering potential than the refractive index, and the loss function (eq.~\ref{eq:loss_fn}) and its gradient are
\begin{align*}
    \mathcal{L}(V) &= \frac{1}{N} \frac{1}{2 M} \sum_{i=1}^M \sum_{j=1}^N \left \vert \left(F^{(i)} \tilde{V} \right)_j - \tilde{\Psi}^{(i)}_j \right \vert^2\\
    \nabla_{\tilde{V}} \mathcal{L} &= \frac{1}{N} \frac{1}{M} \sum_i \left(F^{(i)} \right)^\dag \left(F^{(i)} \tilde{V} - \tilde{\vec{\Psi}}^{(i)} \right)
\end{align*}
where $\tilde{\vec{\Psi}}^{(i)}$ is the scattered field or the Rytov phase depending on the approximation, $F^{(i)}$ is the forward model linear operator for the $i$th angle which connects the 3D Fourier transform of the scattering potential, $\tilde{V}$, to $\tilde{\vec{\Psi}}^{(i)}$. The second factor of $1/N$ converts this to the loss in real-space accounting for the Fourier transform. Since the forward model is linear, a Lipschitz constant for the loss function is proportional to the largest eigenvalue of $F^t F$ which can be computed using a singular value decomposition or the power iteration algorithm.

\subsection{Multi-slice models\label{section:multislice_models}}
Suppose that we can rewrite the Helmholtz equation (eq.~\ref{eq:helmholtz}) by reparameterizing the electric field as $\Psi$ and rewriting the differential operator as the sum of two terms, where $A$ describes the effect of the background refractive index $n_o$ and $B$ captures the effect of spatially varying refractive index perturbations,
\begin{align}
    \partial_z \Psi &= \left(A + B \right) \Psi. \label{eq:helmholtz_separated}
\end{align}
Formally the solution is a path-ordered exponential, but for small enough $\delta z$,
\begin{align}
    \Psi(\delta z) &\approx \exp \left[(A + B) \delta z \right] \Psi(0)\\
    &\approx \exp \left[A \delta z \right] \exp \left[B \delta z \right] \Psi(0). \label{eq:cbh_approx}
\end{align}
For convenience we define $P = \exp \left[A \delta z \right]$ and $Q = \exp \left[B \delta z \right]$. Corrections to eq.~\ref{eq:cbh_approx} are given by the Baker-Campbell-Hausdorff formula, and the first correction is proportional to the commutator $[A, B]$. By construction, this vanishes when $n = n_o$, and we expect it and higher order commutators to be small when the refractive index perturbation is small.

We can propagate an initial field through a volume by discretizing it into layers of thickness $\delta z$ and iteratively applying eq.~\ref{eq:cbh_approx}
\begin{align}
    \Psi &= F (PQ)^{(k-1)} ... (PQ)^{(0)} \Psi^{(0)}
\end{align}
Here $F$ describes model operations beyond the final refractive index plane, $\Psi^{(m)}$ is the intermediate field before the $m$th voxel, and $\Psi$ is the detected field.

For models considered here, $Q$ is local in the sense that the only dependence on $n(z_l)$ is in $Q^{(l)}$, and thus 
\begin{align}
    \frac{\partial  \Psi_a}{\partial n_b(z_l)} &=  \left(W^{(l + 1)} P^{(l)} \frac{\partial  Q^{(l)}}{\partial n_b(z_l)} \Psi^{(l)}\right)_a \label{eq:de_dn}\\
    W^{(l+1)} &= F (PQ)^{(k-1)} ... (PQ)^{(l + 1)}, \label{eq:w_def}
\end{align}
where the index $a$ in the first equation indicates the $a$th component of the vector. Typically the structure of $\partial Q$ allow us to simplify this expression. For example, in the BPM $Q$ is diagonal due to the fact it does not mix the field at different $\vec{r}$ positions. For the SSNP, $Q$ is local but mixes the derivative and field at the same position.

\subsection{Beam-propagation model (BPM)\label{section:bpm}}
In the BPM, the Helmholtz equation is put in the form of eq.~\ref{eq:helmholtz_separated} by first making the paraxial approximation~\cite{Kamilov2016}. Here we take $\Psi = E$ and the model is defined by
\begin{align}
    P \Psi &= \ift{\exp \left[i \delta z \sqrt{k^2 - k_x^2 - k_y^2} \right] \times \ft{\Psi}(k_x, k_y)}\\
    Q^{(l)} \Psi &= \exp \left[ik_o \delta z \frac{n(x, y, z_l) - n_o}{\eta} \right] \times \Psi \\
    F \Psi &= \ift{H(k_x, k_y) \times \exp \left[ i\delta z_f \sqrt{k^2 - k_x^2 - k_y^2} \right] \times \ft{\Psi}(k_x, k_y) }
\end{align}
where $H$ is the coherent transfer function, $\delta z_f$ is the final distance the beam propagates, and $k_o = 2 \pi / \lambda$. $\eta$ is the obliquity factor which is taken to be $1$ in most cases, but taking $\eta = 1/\cos \theta$ improves accuracy~\cite{zhuHighfidelityIntensityDiffraction2022}.

Following eqs.~\ref{eq:loss_fn_grad} and \ref{eq:de_dn} the derivatives are
\begin{align}
    \frac{\partial Q^{(l)}_{cd}}{\partial n_b(z_l)} &= \frac{ik_o \delta z}{\eta} \ \exp \left[ik_o\delta z \frac{n_c(z_l)-n_o}{\eta} \right] \times \delta_{cd} \times \delta_{bd}\\
    \frac{\partial \Psi_a}{\partial n_b(z_l)} &= \frac{ik_o \delta z}{\eta} W^{(l+1)}_{ab} \Psi^{(l+1)}_b\\
    \nabla_{n(z_l)} \mathcal{L} &= -\frac{ik_o \delta z}{\eta} \left[\left(W^{(l+1)}\right)^\dag \partial_{\vec{\Psi}^*} \mathcal{L} \right] \odot \left(\Psi^{(l+1)}\right)^*.
\end{align}

\subsection{Split-step non-paraxial model (SSNP)\label{section:ssnp}}
This model is extensively discussed elsewhere~\cite{Lim2019, zhuHighfidelityIntensityDiffraction2022} and we briefly discuss it here for convenience. To rewrite the Helmholtz equation in the form of eq.~\ref{eq:helmholtz_separated} we must work with a vector of the electric field and its first derivative. The physical intuition behind this is as follows. Suppose we know the electric field in a single plane and wish to know its value everywhere. We can decompose the field into Fourier modes, but we cannot distinguish forward and backwards travelling plane waves at the same lateral spatial frequency. We need additional information to untangle these two contributions: e.g. the magnetic field or the axial derivative of the electric field $\partial_z E$. Therefore the propagation operator must act on both the field and its derivative.

In this case the forward model is
\begin{align}
    \Psi &= 
    \begin{pmatrix} 
    E\\
    \partial_z E
    \end{pmatrix}\\
    P \Psi &= 
    \ift{
    \begin{pmatrix}
    \cos k_z \delta z & \frac{1}{k_z} \sin k_z \delta z\\
    - k_z \sin k_z \delta _z & \cos k_z \delta _z
    \end{pmatrix}
    \ft{\Psi}
    }\\
    Q^{(l)} \Psi &=
    \begin{pmatrix}
        1 & 0\\
        k_o^2 \left[ n_o^2 - n^2(x, y, z_l) \right] \delta z & 1
    \end{pmatrix} \Psi\\
    F \Psi &= 
    \ift{
    \begin{pmatrix}
     \frac{1}{2} & \frac{1}{2} \frac{1}{i k_z}   
    \end{pmatrix}
    H(k_x, k_y) \times P(\delta z_f) \Psi
    }.
\end{align}

When we write $Q$ as a matrix following eq.~\ref{eq:de_dn} we combine the structure of the $\vec{r}_\perp$ index and the field/derivative index. Let $i$ and $j$ index the position $\vec{r}_\perp$ and field/derivative respectively and define the composite index $c = 2 i + j$. Following eqs.~\ref{eq:loss_fn_grad}, \ref{eq:de_dn}, and \ref{eq:w_def}, the derivatives are
\begin{align}
    \frac{\partial Q_{cd}}{\partial n_b} &= -2 k_o^2 \ \delta z \ n \times \delta_{c, 2b + 1} \ \delta_{d, 2b}\\
    \frac{\partial E_a}{\partial n_b(z_l)} &=  -2 k_o^2 \ \delta z \ n_b \times Y^{(l + 1)}_{a, 2b+1} \Psi_{2b}^{(l)}\\
    \frac{\partial \mathcal{L}}{\partial n_b(z_l)} &= -k_o^2 \ \delta z  \sum_a \left(Y_{a, 2b + 1}^{(l+1)} \right)^* D_a \left(\Psi_{2b}^{(l)} \right)^* n_b^*(z_l)\\
    \nabla_n \mathcal{L} &=  -2k_o^2 \ \delta z \left[ \left(\hat{Y}^{(l+1)} \right)^\dag \partial_{\vec{\Psi}^*} \mathcal{L} \right] \odot n^*(z_l) \odot \left(\vec{E}^{(l)}\right)^*.
\end{align}
where $Y^{(l+1)} = W^{(l+1)} P^{(l)}$ and in the last line we use $\Psi_{2b} = E_b$ and define $\hat{Y}$, a modified version of $Y$ which retains only the derivative parts of the second index.

\end{document}